\documentclass{emulateapj}

\usepackage{verbatim}
\usepackage{amsmath}
\usepackage{subfigure}

\def\M{{\cal M}}

\newcommand{\simgt}{\,\hbox{\lower0.6ex\hbox{$\sim$}\llap{\raise0.6ex\hbox{$>$}}}\,}
\newcommand{\simlt}{\,\hbox{\lower0.6ex\hbox{$\sim$}\llap{\raise0.6ex\hbox{$<$}}}\,}
\newcommand{\degree}{\ensuremath{^\circ}}

\begin{document}

\title{Cold Fronts and Gas Sloshing in Galaxy Clusters with Anisotropic Thermal Conduction}

\author{J. A. ZuHone\altaffilmark{1}, M. Markevitch\altaffilmark{1}, M. Ruszkowski\altaffilmark{2,3}, 
  D. Lee\altaffilmark{4}}

\altaffiltext{1}{Astrophysics Science Division, Laboratory for High Energy Astrophysics, Code 662, NASA/Goddard Space Flight Center, Greenbelt, MD 20771}
\altaffiltext{2}{Department of Astronomy, University of Michigan, 500 Church Street, Ann Arbor, MI 48109}
\altaffiltext{3}{The Michigan Center for Theoretical Physics, 3444 Randall Lab, 450 Church St, Ann Arbor, MI 48109}
\altaffiltext{4}{The Flash Center for Computational Science, The University of Chicago, 5747 S. Ellis, Chicago, IL 60637, USA}

\begin{abstract}
Cold fronts in cluster cool cores should be erased on short
timescales by thermal conduction, unless protected by
magnetic fields that are ``draped'' parallel to the front
surfaces, suppressing conduction perpendicular to the sloshing fronts. We
present a series of MHD simulations of cold front formation in the
core of a galaxy cluster with anisotropic thermal
conduction, exploring a parameter space of conduction strengths
parallel and perpendicular to the field lines. Including conduction has a strong effect on the
temperature distribution of the core and the appearance of the cold
fronts. Though magnetic field lines are draping parallel to
the front surfaces, preventing conduction directly across them, the
temperature jumps across the fronts are nevertheless reduced. The geometry
of the field is such that the cold gas below the front surfaces can be
connected to hotter regions outside via field lines along directions perpendicular to the plane of the sloshing
motions and along sections of the front which are not perfectly draped. This results in the
heating of this gas below the front on a timescale of a Gyr, but the
sharpness of the density and temperature jumps may be nevertheless preserved.  By modifying the gas
density distribution below the front, conduction may indirectly aid in
suppressing Kelvin-Helmholtz instabilities. If conduction along the
field lines is unsuppressed, we find that the characteristic sharp
jumps seen in {\it Chandra} observations of cold front clusters do not form; therefore,
the presence of of cold fronts in {\it hot} clusters is in
contradiction with our simulations with full Spitzer conduction. This
suggests that the presence of cold fronts in hot clusters could
be used to place upper limits on conduction in the {\it
  bulk} of the ICM. Finally, the combination of sloshing and
anisotropic thermal conduction can result in a larger flux of heat to
the core than either process in isolation. While still not sufficient to prevent a cooling
catastrophe in the very central ($r \sim$~5~kpc) regions of the cool
core (where something else is required, such as AGN feedback), it reduces significantly
the mass of gas that experiences a cooling catastrophe outside those
small radii.
\end{abstract}


\section{Introduction\label{sec:intro}}

X-ray observations of galaxy clusters often show a bright, dense,
centrally peaked core comprised of the intracluster medium (ICM) much
cooler than the cluster average. The X-ray radiative cooling time,
inversely proportional to the square of the gas density, is very short
in these dense cores, on the order of several hundred Myr in the very
central regions \citep{sar86}. This gave rise to an early ``cooling
flow'' scenario, in which the gas cools catastrophically in the
cluster center and is continuously replaced by the inflow of the
surrounding cooling gas. However, {\it Chandra} and {\it XMM-Newton} have failed
to detect the implied large quantities of very cool gas in the cores;
in particular, the characteristic emission lines such as Fe XVII
emitted by the plasma with $T<1$ keV are not observed \citep{pet06}. Since X-ray cooling is directly observed, a mechanism is
required to heat the ICM in the core to compensate for the cooling.
The two most studied heating mechanisms are feedback from active
galactic nuclei (AGN) and thermal conduction. In this paper, we will
be discussing the latter mechanism in connection with the phenomenon
of gas sloshing.

The role of thermal conduction in the context of heating the cool
cores of clusters of galaxies has been studied by many authors
\citep[see, e.g.,][]{bin81, nar01, rus02, zak03, guo08}. Since in the ICM the Larmor
radius of the electrons spiraling along the $B \sim 0.1-1$~$\mu$G magnetic field lines is
many orders of magnitude smaller than the electron mean free path, the
conduction is strongly anisotropic, occurring essentially exclusively
along the field lines. Therefore, most of these works
assumed an isotropically tangled magnetic field geometry and
paramaterized the heat flux as an effective isotropic thermal
conductivity given as a fraction, $f_{\rm Sp}$, of the ideal (field-free) \citet{spi62}
heat flux. 

However, theoretical and numerical work over the last decade
has revealed that the stability and thermal properties of the ICM are
possibly more complicated than previously thought if anisotropic
thermal conduction is modelled explicitly in the MHD regime. Early analysis by
\citet{bal00,bal01} showed that when magnetized atmospheres with
anisotropic thermal conduction are considered, their stability depends
on the {\it temperature} gradient rather than the entropy gradient. He
demonstrated that in regions where the temperature increases in the
direction of gravity, the plasma is susceptible to the magnetothermal
instability (hereafter MTI).

Further analytical work by \citet{qua08} and numerical work by
\citet{par08a} and \citet{par08b,par09} has shown that the cluster atmosphere is unstable even
when the temperature decreases in the direction of gravity, as in the
cluster cool cores (heat-flux
buoyancy instability; HBI). The effect of the HBI is to realign the
magnetic fields azimuthally
$(\hat{\textbf{\emph{b}}}\cdot\hat{\textbf{\emph{{r}}}} \approx 0)$,
and thereby significantly reduce the effective radial heat flux
$Q_r$. This renders thermal conduction almost useless against
preventing a ``cooling catastrophe.'' Subsequent work has shown that
relatively low levels of turbulence can rearrange the field lines in a
more isotropic configuration, permitting heat conduction to the core
\citep{par10,rus10,rus11a}. More recently, and of direct relevance to
this work, \citet{lec12} simulated the dynamics of Rayleigh-Taylor
(RT) stable and unstable contact discontinuities with anisotropic
thermal conduction. Of particular relevance is their finding that the
temperature gradients of RT-stable contact discontinuities (such as
sloshing cold fronts) can be smoothed by anisotropic thermal
conduction if the field lines are at least partially misaligned with
the surface of the discontinuity. The smoothing of the temperature
gradient of the RT-stable contact discontinuities in these simulations was limited by the HBI,
which reoriented magnetic field lines initially perpendicular to the front
surface in a more parallel direction, suppressing conduction across
the surface at later times. 

So far, the numerical simulations of the evolution of the cluster atmospheres with anisotropic thermal conduction have mostly been
limited to isolated, spherically symmetric clusters initially in
hydrostatic equilibrium. These works have been essential in order to
determine the character of the instabilities that are present when
such physical effects are taken into account. A recent exception to this is the paper by
\citet{rus11b}, which simulated the formation of a galaxy cluster from
cosmological initial conditions. 

The previous works assumed that the heat conduction along the field
lines is at the Spitzer value. However, the
effective conductivity along the field lines may be reduced due to strong curvature of the field lines and stochastic
magnetic mirrors on small scales \citep[e.g.,][]{cha99,mal01,nar01}. 
\citet{mar03b} attempted to constrain heat conduction in the ICM from
the observed temperature differences in the merging, non-cool-core
cluster Abell 754. They determined that the effective conduction in
the direction of the observed temperature gradients is an order of magnitude less than
Spitzer, which, under the assumption of an isotropically tangled
field, imples suppression along the field lines. Finally, \citet{sch05} suggested that plasma microscale instabilities may
be very important in the high-$\beta$ plasma of the ICM, and
consequently its thermal conductivity may be many orders of magnitude
smaller than the Spitzer value \citep{sch08}. 

In this work we consider a scenario where the initial cluster core is relaxed
but undergoes a period of gas ``sloshing.'' Many cool-core systems
exhibit edges in X-ray surface brightness approximately concentric
with respect to brightness peak of the cluster
\citep[e.g.,][]{maz01,mar01,MVF03}. X-ray spectra of these regions
have revealed that in most cases the brighter (and therefore denser)
side of the edge is the colder side, and hence these jumps in gas
density have been dubbed ``cold fronts'' \citep[for a detailed review
see][]{MV07}. These features have been observed in the majority of
cool-core systems \citep[][]{ghi10}. The presence of one or more cold fronts is seen as an
indication of the motion of gas in a cluster. For relaxed
systems, the primarily spiral-shaped cold fronts are believed to
arise from gas sloshing in the deep dark matter-dominated potential
well. These motions are initiated when the low-entropy, cool gas of
the core is displaced from the bottom of the dark matter potential
well, either by gravitational perturbations from infalling subclusters
\citep[][]{AM06} or by an interaction with a shock front
\citep[][]{chu03}. 

In an earlier paper, \citet[][hereafter ZMJ10]{zuh10}, we
have investigated the possible flow of heat to the cluster cool core as a result
of the sloshing motions bringing hot gas from larger radii into contact with
the cool gas of the core. We determined that if the gas viscosity is
small, significant mixing between the hot and cold gas will result, raising the entropy of the
core, though this heat flow was found to be insufficient to completely
offset the effects of radiative cooling. In that work, we did not
include thermal conduction, but only mixing of gases with different temperatures.

ZMJ10 modeled the ICM as a (possibly viscous) unmagnetized
fluid. However, the ICM is known to be magnetized, and this fact has
important considerations for heat transport. For example, it has been suggested
that the narrow width of cold fronts in galaxy clusters can be used to
constrain the effectiveness of heat conduction in the ICM
\citep[e.g.,][]{ett00,xia07}. The sharp temperature gradient of the
cold front should be smoothed out on short timescales if conduction is
not suppressed across the front surface. Such suppression could be
provided by magnetic field lines oriented parallel to the cold front
surfaces, a situation believed to naturally arise due to the effects
of magnetic draping and shear amplification \citep{asa04,asa05,lyu06,asa07,kes10}. In a
previous paper \citep[][hereafter ZML11]{zuh11}, we performed
simulations of gas sloshing with magnetic fields, showing that such
shear amplification and draping does occur around cold fronts in
cluster cores. These simulations also showed that magnetic fields have
similar effects as viscosity on the ICM, inhibiting the mixing of the
hot and cold flows and reducing the increase of core entropy due to sloshing.

In this work, we build upon our earlier simulations (ZMJ10, ZML11) by including the effects of
anistropic thermal conduction in a simulation of gas sloshing in a
cool-core galaxy cluster. We will show that the magnetic fields
draped across the cold fronts indeed inhibit the smoothing out of the
temperature gradient, but only to a degree. The crucial points to note
are a) due to fluid instabilities the magnetic field lines may
not be oriented perfectly azimuthal with the front and b) cold fronts
are not completely enclosed surfaces--there are still field lines that
connect the cold side of the cold fronts to regions of
higher temperature in the cluster atmosphere over short
distances. Therefore, if thermal conduction is important, it may provide a more efficient way
than gas mixing to transfer heat between the hot and cold flows characteristic of
sloshing motions. 

This paper is organized as follows: in Section 2 we
describe the simulations and the code. In Section 3 we
describe the effects of anisotropic conduction and gas sloshing on the
thermal state of the cluster core for different configurations of the
magnetic field and various prescriptions for thermal
conduction. Finally, in Section 4 we summarize our
results and discuss future developments of this work. We assume a flat
$\Lambda$CDM cosmology with $h = 0.7$, $\Omega_{\rm m} = 0.3$, and
$\Omega_b = 0.02h^{-2}$.  

\section{Method\label{sec:method}}

In our simulations, we solve the following set of MHD equations (in Gaussian units):
\begin{eqnarray}
\frac{\partial{\rho}}{\partial{t}} + \nabla \cdot (\rho{\bf v}) &=& 0 \\
\frac{\partial{(\rho{\bf v})}}{\partial{t}} + \nabla \cdot \left(\rho{\bf vv} - \frac{\bf BB}{4\pi}\right) + \nabla{p} &=& \rho{\bf g} \\
\frac{\partial{E}}{\partial{t}} + \nabla \cdot \left[{\bf v}(E+p) - \frac{{\bf
  B}({\bf v \cdot B})}{4\pi}\right] &=& \rho{\bf g \cdot v} - \nabla
\cdot {\bf Q} \\
\nonumber &-& n_en_H\Lambda(T,Z) \\
\frac{\partial{\bf B}}{\partial{t}} + \nabla \cdot ({\bf vB} - {\bf Bv}) &=& 0
\end{eqnarray}
where
\begin{eqnarray}
p = p_{\rm th} + \frac{B^2}{8\pi} \\
E = \frac{\rho{v^2}}{2} + \epsilon + \frac{B^2}{8\pi}
\end{eqnarray}
where $p_{\rm th}$ is the gas pressure, $\epsilon$ is the gas internal
energy per unit volume, and $\bf{B}$ is the magnetic field strength. For all our simulations, we assume an ideal
equation of state with $\gamma = 5/3$ and primordial abundances with
molecular weight $\mu = 0.6$. 

We performed our simulations using FLASH 3.3, a parallel
hydrodynamics/$N$-body astrophysical simulation code developed at the
FLASH Center for Computational Science at the University of
Chicago\citep{fry00,dub09}. FLASH 3.3 solves the equations of
magnetohydrodynamics using a directionally unsplit staggered mesh
algorithm \citep[USM;][]{lee09}. The USM algorithm used in FLASH 3.3
is based on a finite-volume, high-order Godunov scheme combined with a
constrained transport method (CT), which guarantees that the evolved
magnetic field satisfies the divergence-free condition
\citep{eva88}. In our simulations, the order of the USM algorithm
corresponds to the Piecewise-Parabolic Method (PPM) of \citet{col84},
which is ideally suited for capturing shocks and contact
discontinuties (such as the cold fronts that appear in our
simulations).  

The heat flux due to thermal conduction is given as:
\begin{equation}
{\bf Q} = -\kappa_{\rm Sp}\left[f_\perp\left({\bf I} - \hat{\textbf{\emph{b}}}\hat{\textbf{\emph{b}}}\right)+f_\parallel \hat{\textbf{\emph{b}}}\hat{\textbf{\emph{b}}}\right] \cdot \nabla{T}
\end{equation}
where we chose to represent the conductivity coefficient as a product
of the Spitzer conductivity $\kappa_{\rm Sp}$ and the fractions $f_\parallel$ and
$f_\perp$ of Spitzer conductivity parallel and
perpendicular to the field line, respectively. $\hat{\textbf{\emph{b}}}\hat{\textbf{\emph{b}}}$ and ${\bf
  I} - \hat{\textbf{\emph{b}}}\hat{\textbf{\emph{b}}}$ are projection operators that single out
the components of the temperature gradient $\nabla{T}$ parallel and
perpendicular to the magnetic field line, with $\hat{\textbf{\emph{b}}}$ the unit
vector in the direction of the local magnetic field. Most works assume
$f_\parallel = 1$ and $f_\perp = 0$ in the conditions of the ICM
(e.g., heat conducted strictly along the field lines without suppression), due
to the large mean free path for collisions compared to the Larmor
radius of the electron. We assume this prescription as the default,
but also investigate alternative scenarios, represented by simulations with different values of $f_\parallel$ and $f_\perp$. For example, if the magnetic field is
strongly curved on small scales, the effective conductivity along the field lines may be smaller than the Spitzer value, resulting in a smaller $f_\parallel$. 

Alternatively, if a fraction of the magnetic field lines draped along the cold front surfaces reconnect with lines below these surfaces, a small amount of heat conduction
perpendicular to the cold front surfaces may result. Our ideal MHD simulations do not model reconnection explicitly, as non-ideal terms including resistivity are not included (there is a small numerical resistivity associated with the finite resolution of the simulation, but this effective resistivity is negligible, see Section \ref{sec:}). However, we may account for the conduction that may occur across cold front surfaces in an approximate way by modeling it as a non-zero $f_\perp$, with the same temperature dependence as the parallel conduction. Since we are not modeling the reconnecting field lines and the conduction along them explicitly, our simple model is intended to provide only an order of magnitude estimate of this possible effect. 

It is important to distinguish our "perpendicular" conduction across the cold front surfaces from conduction perpendicular to the field lines resulting from collisions between electrons. Because the Larmor radius is many orders of magnitude smaller than the electron mean free path, this perpendicular conduction is effectively zero for the conditions in galaxy clusters. Additionally, the temperature dependence of this conduction would be $\propto T^{-1/2}$ instead of the $\propto T^{5/2}$ dependence for conduction along the field lines \citep{bra65}. In contrast, since our "perpendicular" conduction is intended to account for conduction that occurs parallel to reconnecting field lines, it has the same temperature dependence as the parallel conduction. 

Following \citet{cow77}, we included the
effect of conduction saturation whenever the characteristic lengthscale
associated with the temperature gradient is smaller than the mean free path,
though in the bulk of the ICM this effect is not significant. We
implemented anisotropic conduction following the approach of
\citet{sha07}. More specifically, we applied a monotonized central
(MC) limiter to the conductive fluxes. This method ensures that
anisotropic conduction does not lead to negative temperatures in the presence of
steep temperature gradients. A verification test of our implementation
of the anisotropic conduction module in FLASH 3.3 was given in
\citet{rus11b}. 

We want to examine the effects of anisotropic thermal conduction
in a sloshing galaxy cluster core in isolation (in particular, the
effect on the cold fronts), and so for most of our
simulations, we do not include the effects of radiative
cooling. We then include radiative cooling to
determine the effect of sloshing with anisotropic conduction on the
thermal state of the cluster core. The X-ray cooling
function is determined using the \citet{MKL95} model with a uniform
metallicity of $Z = 0.3Z_\sun$.

The gravitational potential on the grid is set up as the sum of two
``rigid bodies'' corresponding to the contributions to the potential
from both clusters. This approach to modeling the potential is
used for simplicity and speed over solving the Poisson equation for
the matter distribution, and is an adequate approximation for our
purposes. It is the same approach that we used in ZML11, and is
justified in \citet{rod11}. 

\begin{table*}[thdp]
\caption{Simulation Parameter Space\label{tab:Params}}
\begin{center}
\begin{tabular}{cccccc}
\hline
\hline
Simulation & Relaxed/Sloshing & Magnetic Field & $f_{\parallel}$ & $f_{\perp}$ &
X-ray Cooling? \\
\hline
S1 & Sloshing & Tangled & 0.0 & 0.0 & NO \\
S2 & Sloshing & Azimuthal & 0.0 & 0.0 & NO \\
SC1 & Sloshing & Tangled & 1.0 & 0.0 & NO \\
SC2 & Sloshing & Azimuthal & 1.0 & 0.0 & NO \\
SC3 & Sloshing & Tangled & 0.1 & 0.0 & NO \\
SC4 & Sloshing & Tangled & 0.1 & 0.01 & NO \\
RX & Relaxed & Tangled & 0.0 & 0.0 & YES \\ 
RCX & Relaxed & Tangled & 1.0 & 0.0 & YES \\
SX & Sloshing & Tangled & 0.0 & 0.0 & YES \\
SCX1  & Sloshing & Tangled & 1.0 & 0.0 & YES \\
SCX2 & Sloshing & Azimuthal & 1.0 & 0.0 & YES \\
SXNoB & Sloshing & No Fields & 0.0 & 0.0 & YES \\
\hline
\end{tabular}
\end{center}
\end{table*}

We refer the reader to ZML11 for the full details of our physical
setup, but will provide the following short description. Our
simulations consist of a massive ($M \approx 1.5
\times 10^{15} \M_\sun$, $T_X \sim 8$~keV), cool-core cluster, initially in hydrostatic equilibrium, merging
with a small (mass ratio $R$ = 5) gasless subcluster, which sets off
the sloshing of the cool core. We begin the simulation at a point in
time when the cluster centers have a mutual separation of $d$ =
3~Mpc and an impact parameter $b$ = 500~kpc. The initial velocities of
the subclusters are set up assuming that the total kinetic energy of
the system is set to half of its total potential energy. For all of the simulations, we set up the main cluster within a cubical computational domain of width $L = 2.4$~Mpc on a side, with a finest cell size on our AMR grid of $\Delta{x} = 2.34$~kpc. 

The magnetic field of the cluster is set up in an identical way to the
simulations of ZML11, following the approach of \citet{rus07, rus08} and
\citet{rus10}. A random magnetic field $\tilde{\bf B}({\bf k})$ is set up in $\bf{k}$-space on a uniform grid using independent normal random deviates for the real and imaginary components of the field. Thus, the components of the complex magnetic field in $\bf{k}$-space are set up such that
\begin{eqnarray}
\tilde{B_x}(\bf{k}) &=& B_1[N(u_1) + iN(u_2)] \\
\tilde{B_y}(\bf{k}) &=& B_2[N(u_3) + iN(u_4)] \\
\tilde{B_z}(\bf{k}) &=& B_3[N(u_5) + iN(u_6)] 
\end{eqnarray}
where $N(u)$ is a function of the uniformly distributed random variable $u$ that returns Gaussian-distributed random values, and the values $B_i$ are field amplitudes. We adopt a dependence of the magnetic field amplitude $B(k)$ on the wavenumber $|\bf{k}|$ similar to (but not the same as) \citet{rus07} and \citet{rus10}:
\begin{equation}
B(k) \propto k^{-11/6}{\rm exp}[-(k/k_0)^2]{\rm exp}[-k_1/k]
\end{equation}
where $k_0$ and $k_1$ control the exponential cutoff terms in the
magnetic energy spectra. The cutoff at high wavenumber $k_0$ roughly
corresponds to the coherence length of the magnetic field $k_0 =
2\pi/\lambda_0 = 2\pi/(43~{\rm kpc})$, observed in some cluster cores.
The cutoff at low wavenumber
$k_1 = 2\pi/\lambda_1$ roughly corresponds to $\lambda_1 \approx
r_{500}/2 \approx$~500~kpc, which is held fixed for all of our
simulations. This field spectrum corresponds to a Kolmogorov shape for
the energy spectrum $(P_B(k) \propto k^{-5/3})$ with cutoffs at large
and small linear scales. This field is then Fourier transformed to yield ${\bf B}({\bf x})$, which is rescaled to have an average value
of $\sqrt{8\pi{p}/\beta}$ to yield a field with a pressure that
is proportional to the gas pressure, i.e., to have a spatially uniform $\beta$
for the initial field. We set the initial $\beta = 400$. This procedure ensures that there will be significant power at all length scales between $\lambda_1$ and $\lambda_0$, which will result in a field that is significantly tangled down to the scale of $\lambda_0$.

As mentioned above, previous simulation works have predicted that due to the action
of the HBI, an azimuthal magnetic field should arise in the cluster
core, in the absence of turbulence driven by galaxy motions or
AGN. For this purpose, we also include a simulation that has a purely
azimuthal initial field similar to that used in \citet{bog09}, described in spherical coordinates by
\begin{eqnarray}
B_r &=& 0 \\ \label{eqn:Br}
B_\theta &=& 2B_0\sin{\theta}\cos{2\phi} \\ \label{eqn:Btheta}
B_\phi &=& -B_0\sin{2\phi}\sin{2\theta} \label{eqn:Bphi}
\end{eqnarray}
Its magnitude was then scaled by the gas pressure with $\beta =
400$. 

\begin{figure*}
\begin{center}
\plotone{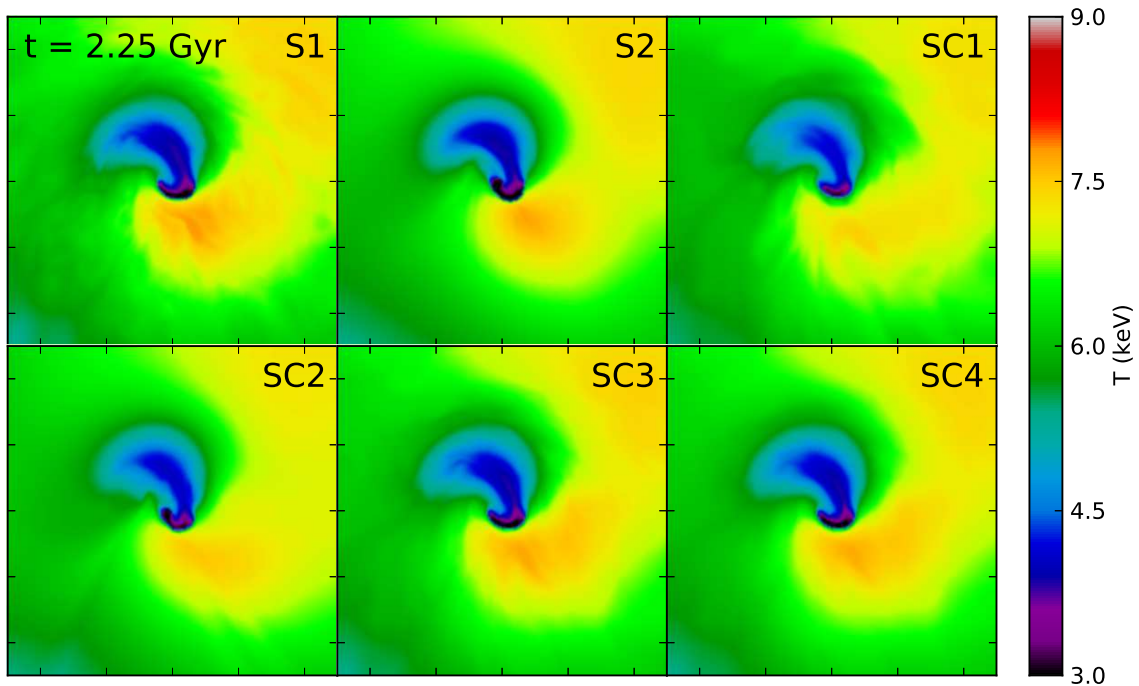}
\caption{Slices through the center of the domain in the $x-y$ plane of
  the temperature for the sloshing simulations with and without
  conduction for the epoch $t$ = 2.25~Gyr. Each panel is 500~kpc on a
  side. The colorscale is temperature in keV. Major tick marks
  indicate 100~kpc distances.\label{fig:t2.25}}
\end{center}
\end{figure*}

Finally, for both configurations, a field consistent with $\nabla
\cdot {\bf B} = 0$ is generated by performing the Fourier transform
and then by ``cleaning'' the field of
divergence terms in Fourier space via the the projection
\begin{equation}
\tilde{\bf B}'({\bf k}) = (\bf{I}-\bf{\hat{k}\hat{k}})\tilde{\bf B}({\bf k})
\end{equation}
where $\hat{\bf{k}}$ is the unit vector in $\bf{k}$-space, and {\bf I}
is the identity operator. Note that this operation does not change the
power spectrum of the magnetic field fluctuations. Upon transformation back to real space, the magnetic field satisfies $\nabla \cdot {\bf B} = 0$. This field is then interpolated onto our AMR grid in such a way that the condition $\nabla \cdot {\bf B} = 0$ is maintained.

Because our sloshing simulations begin with the two clusters at a
distance, it takes a period of time ($\sim$2~Gyr) for the
influence of the subcluster to perturb the cool core of the main
cluster and initiate sloshing. If we had began the simulation with
anisotropic thermal conduction ``switched on'', it would alter the
thermal state of the core and the HBI and MTI would alter the cluster
magnetic field spatial configuration long before sloshing had begun, since the growth times of
these instabilities and the conduction timescale for the core are on
the order of $\sim$~a few $10^8$~yr. Worse yet, if we began the simulation with
radiative cooling switched on, the cool core would cool rapidly (also
on a timescale of $\sim$~a few $10^8$~yr) and we would have a cooling catastrophe
before sloshing started. To avoid these effects, we switch on conduction (and
cooling in the simulations where it is included) at a time $t \sim
2$~Gyr after the beginning of the simulation, shortly after the
subcluster has made its pericentric passage of the cluster core and
right at the beginning of the sloshing period. This enables us to
examine the effects of conduction and sloshing together on a typical cool-core
cluster state. 

An unfortunate side-effect of this evolutionary period
before the core passage is that the magnetic field configuration will
slowly evolve away from its initial configuration. Hence, the
initially isotropically tangled field configuration with
$\langle|\hat{\textbf{\emph{b}}}\cdot\hat{\textbf{\emph{{r}}}}|\rangle = 0.5$ evolves
to a slightly more radial field configuration with
$\langle|\hat{\textbf{\emph{b}}}\cdot\hat{\textbf{\emph{{r}}}}|\rangle
\approx 0.56$, and the initially azimuthal field configuration with
$\langle|\hat{\textbf{\emph{b}}}\cdot\hat{\textbf{\emph{{r}}}}|\rangle
= 0$ also evolves to a slightly more radial field configuration with
$\langle|\hat{\textbf{\emph{b}}}\cdot\hat{\textbf{\emph{{r}}}}|\rangle
\approx 0.35$. This is not a major concern, as the former is not too
far off from an isotropically tangled field configuration and the latter is not
too far from the field configuration of the saturation state of the HBI as seen
in simulations \citep{par09,bog09}. In real clusters, there is always
turbulence and small disturbances from galaxies and DM clumps that
keep the field tangled \citep[][]{par10,rus10,rus11a}.

We will refer to the simulations using the following codes. The letter
``R'' will indicate a cluster that is evolved in isolation, whereas
``S'' will indicate a cluster evolved with a gravitational interaction
causing sloshing. ``C'' will indicate anisotropic thermal conduction
is turned on, and ``X'' will indicate X-ray radiative cooling is
turned on. Simulations that differ in the prescription for conduction
and the initial field line configuration will be delimited by
numbers. Table \ref{tab:Params} lists all simulation runs with their
particular parameters.

\section{Results\label{sec:results}}

\subsection{Brief Outline of the Sloshing Process\label{sec:sloshing_outline}}

We have described the initiation and the evolution of sloshing cold
fronts in previous works (e.g., AM06, ZMJ10, ZML11), the latter
paper including the effects of magnetic fields. We refer the reader to
these papers for the full description, but we will outline the process
in brief here. 

\begin{figure*}
\begin{center}
\plotone{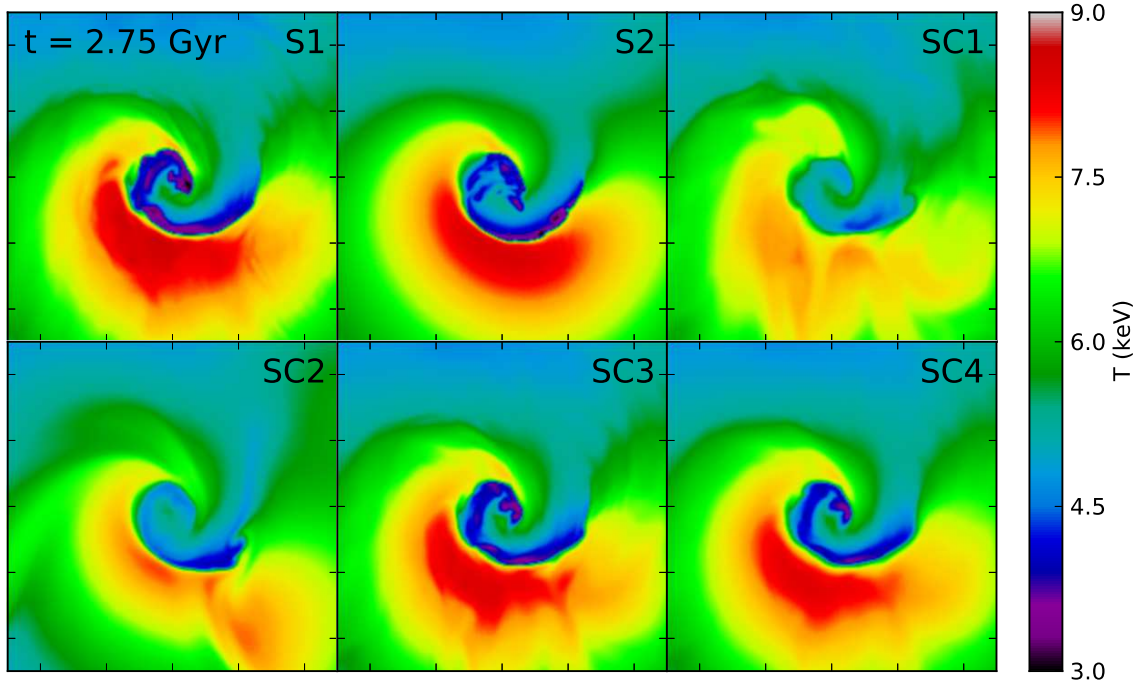}
\caption{Same as Figure \ref{fig:t2.25} but corresponding to the
  epoch $t$ = 2.75~Gyr.\label{fig:t2.75}}
\end{center}
\end{figure*}

As the subcluster approaches the main cluster's core and makes its
passage, the gas and DM peaks of the main cluster feel the same
gravity force toward the subcluster and move together towards
it. However, the gas feels the effect of the ram pressure of the
ambient medium. This fact becomes significant as the gas core is held
back from the core of the dark matter by this pressure. This results
in a separation of the gas and DM peaks of the cluster. As this ram
pressure declines with the decreasing gravitational attraction of the
outgoing subcluster, the gas that was held back by the ram pressure falls
into the DM potential minimum and overshoots it. In addition to the
gravitational disturbance, the wake trailing the subcluster transfers
some of the angular momentum from the subcluster to the core gas and
also helps to push the core gas out of the DM potential well. As
the cool gas from the core climbs out of the potential minimum, it
expands adiabatically. However, the densest, lowest-entropy gas
quickly begins to sink back towards the potential minimum against the
ram pressure from the surrounding ICM. Once again, as the cool gas
falls into the potential well it overshoots it, and the process
repeats itself on a smaller linear scale. Each time, a contact
discontinuity (``cold front'') is produced. Due to the angular
momentum transferred from the subcluster by the wake, these fronts
have a spiral-shaped structure.  As we noted in ZML11, this
general outline is not affected significantly by the strength of the magnetic field or its spatial configuration, so the
features seen in density and temperature maps are essentially the same
between our two reference (e.g., without conduction) simulations, {\it
  S1} and {\it S2} (the {\it S1} simulation corresponds to the {\it Beta400} simulation in ZML11), and the runs
without magnetic fields in ZMJ10. When speaking
of the orientation of the temperature gradient across the cold fronts,
we shall refer to the hotter, less dense gas as the side ``above'' the
front and the colder, denser gas ``below'' the front, consistent with
the direction of the gravitational acceleration. 

\subsection{The Effect of Conduction on Cold Fronts\label{sec:cold_fronts}}

\begin{figure*}
\begin{center}
\plotone{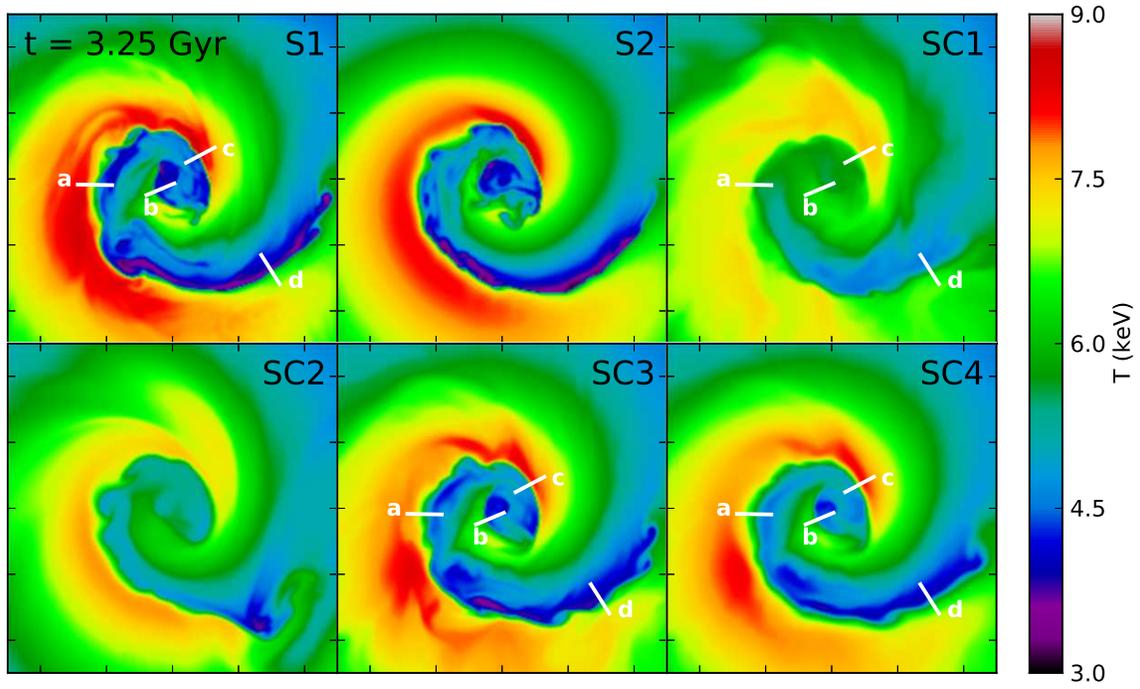}
\caption{Same as Figure \ref{fig:t2.25} but corresponding to the
  epoch $t$ = 3.25~Gyr. The profiles marked with letters correspond to those
  plotted in Figures \ref{fig:temperature_jumps} and \ref{fig:density_jumps}.\label{fig:t3.25}}
\end{center}
\end{figure*}

\begin{figure*}
\begin{center}
\plotone{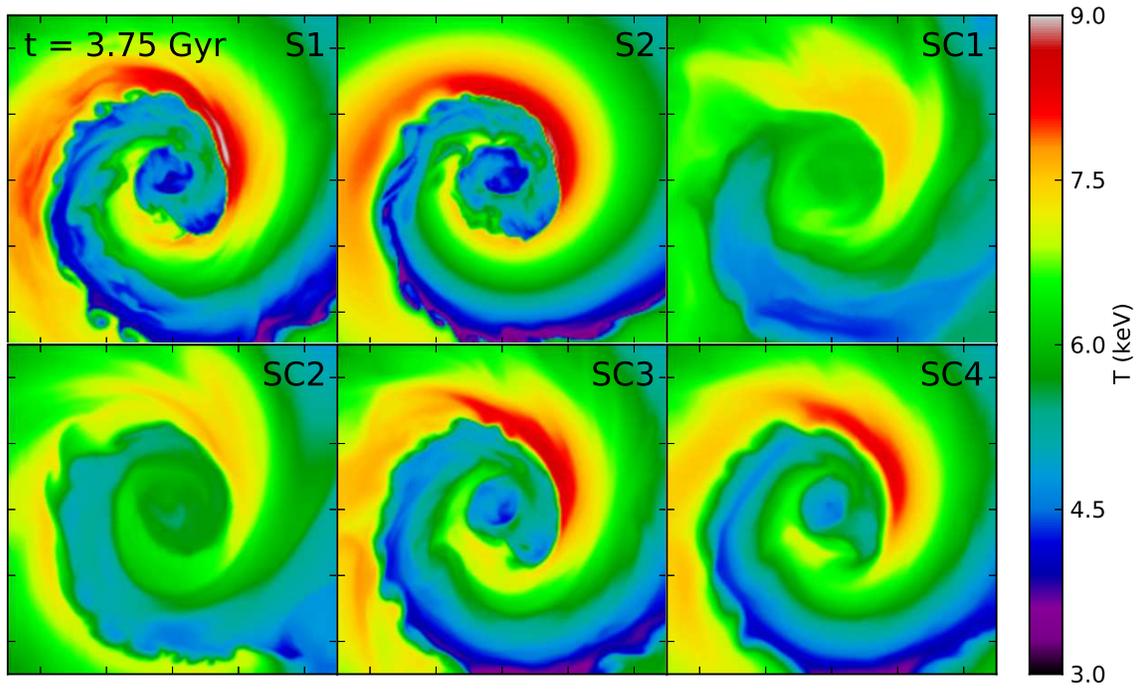}
\caption{Same as Figure \ref{fig:t2.25} but corresponding to the
  epoch $t$ = 3.75~Gyr.\label{fig:t3.75}}
\end{center}
\end{figure*}

Figures \ref{fig:t2.25} through \ref{fig:t3.75} show
slices through the center of the cluster of the gas temperature for several different epochs,
for our simulations with and without anisotropic conduction (but without radiative cooling). We
will first compare runs with no conduction to runs with full Spitzer
conduction along the field lines ($f_\parallel = 1, f_\perp = 0$). Our
runs {\it S1} (top left panels) and {\it S2} (top
center panels) correspond to sloshing simulations with initially
tangled fields and initially azimuthal fields, respectively. Likewise, our runs {\it SC1} (top right
panels) and {\it SC2} (bottom left panels) correspond to sloshing simulations with anisotropic heat
conduction, with initially tangled fields and initially azimuthal fields.  

Including conduction has a dramatic effect on the temperature
structure of the cold fronts. The adiabatic simulations without conduction ({\it S1} and
{\it S2}) develop a number of well-defined cold fronts over time,
which maintain temperature differences of $\sim$1.5-2 across the
fronts throughout their evolution. This is not the case for the
simulations with anisotropic unsuppressed Spitzer conduction (simulations {\it SC1}
and {\it SC2}). Over time, the cold gas inside the fronts is heated up, so that the temperature contrasts
across the fronts become smaller, roughly $\sim$1.3-1.5. The changing
thermal state of the gas also changes the shape of the fronts in some
places. In particular, due to the fact that the inner 50~kpc becomes
nearly isothermal, cold fronts do not occur on the smallest
scales, in conflict with observations of cool-core clusters, which
typically show a number of cold fronts at both small and intermediate radii
within the core. The temperature of the core in simulation {\it SC2} increases
more slowly than in simulation {\it SC1}, because the more azimuthally oriented
field lines in the former suppress conduction from larger radii more
efficiently than the more tangled fields of simulation {\it SC1} do.

Figures \ref{fig:t2.25} through \ref{fig:t3.75} also show results for
the simulations with other values for the conduction coefficients
$f_\perp$ and $f_\parallel$. The {\it SC3} simulation has the
same field line configuration as the {\it SC1} simulation, but the
conduction along the lines is suppressed by a factor of 10. If field
lines are strongly curved on very small linear scales, the effective
conduction coefficient parallel to the field lines may be smaller
than the Spitzer value by an order of magnitude due to the presence of stochastic magnetic
mirrors \citep[e.g.,][]{nar01}. We choose $f_\parallel = 0.1$ to
determine the effect of such a reduction. Alternatively, if plasma
microscale instabilities play a significant role, the thermal
conductivity of the ICM may be many orders of magnitude smaller
\citep{sch08}. For all practical purposes, this case is equivalent to
the runs without conduction.
 
On the other hand, if (for example) $\sim$1\% of the magnetic field lines parallel to the fronts
reconnect, the effective fraction of Spitzer conduction perpendicular to the large-scale
direction of the field could be $f_\perp \sim 0.01$. We assume this
value of $f_\perp$ for simulation {\it SC4}, keeping the value of
$f_\parallel = 0.1$ for comparison with simulation {\it SC3}. As
expected, for the {\it SC3} simulation, far less heat is conducted to
the cluster core, and the temperature contrast of the cold fronts with
the surrounding medium is much closer to the case without any
conduction. In the {\it SC4} simulation, the small perpendicular
conductivity, combined with a steep temperature jump across the cold front, has succeeded
over time to smooth out the temperature gradient across the cold fronts.

\begin{figure*}
\begin{center}
\includegraphics[width=0.45\textwidth]{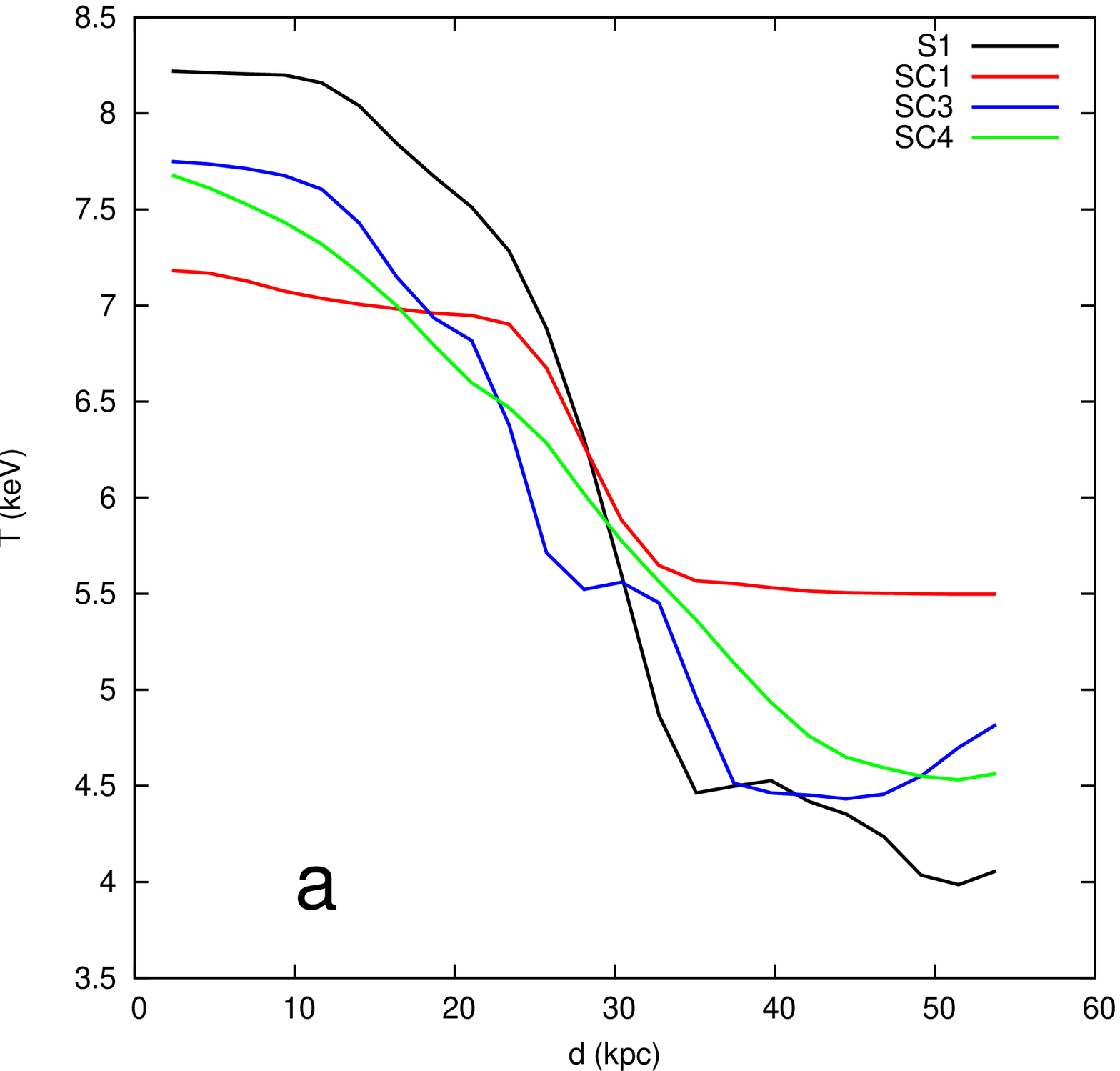}
\includegraphics[width=0.45\textwidth]{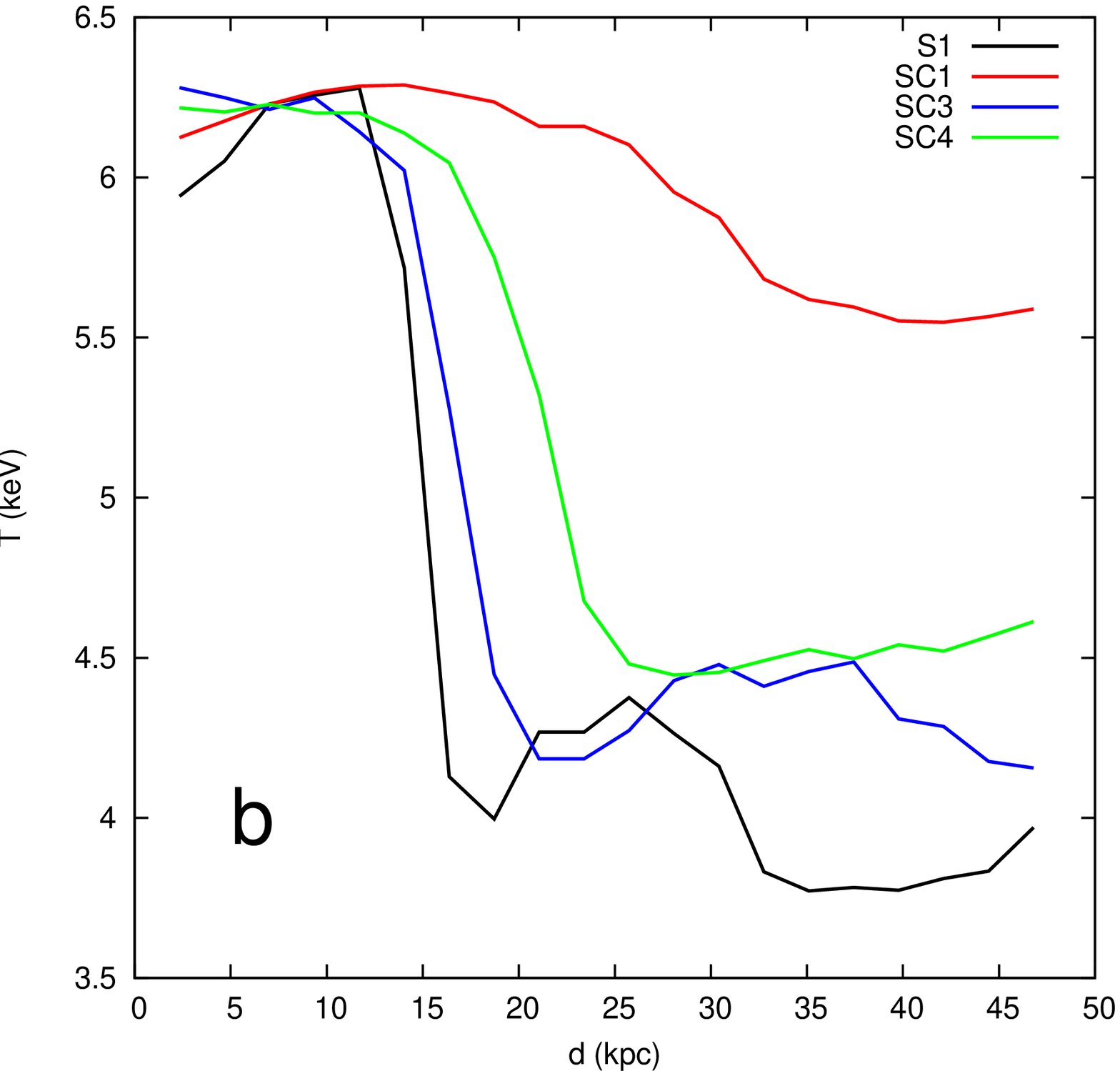}
\par
\includegraphics[width=0.45\textwidth]{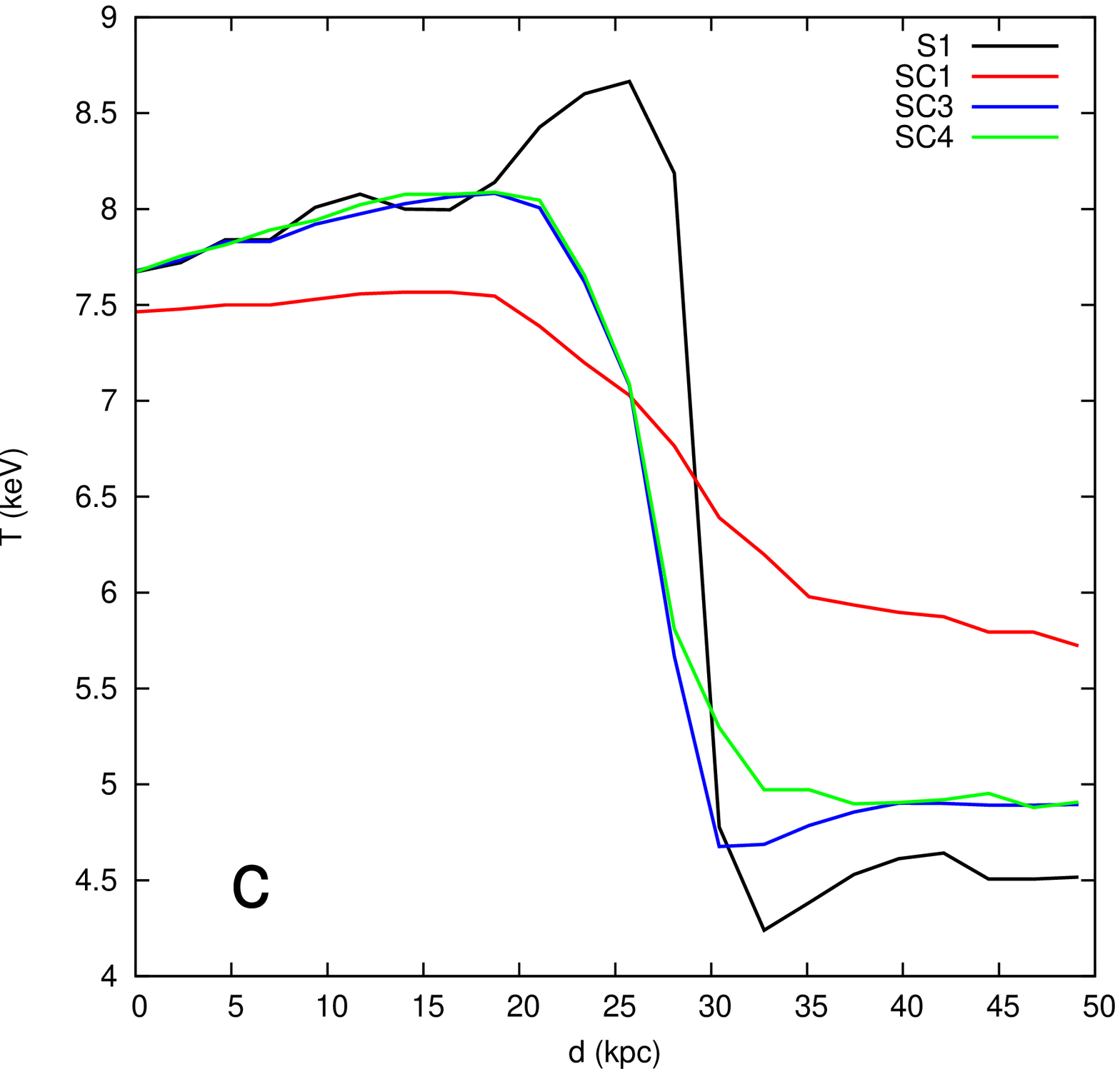}
\includegraphics[width=0.45\textwidth]{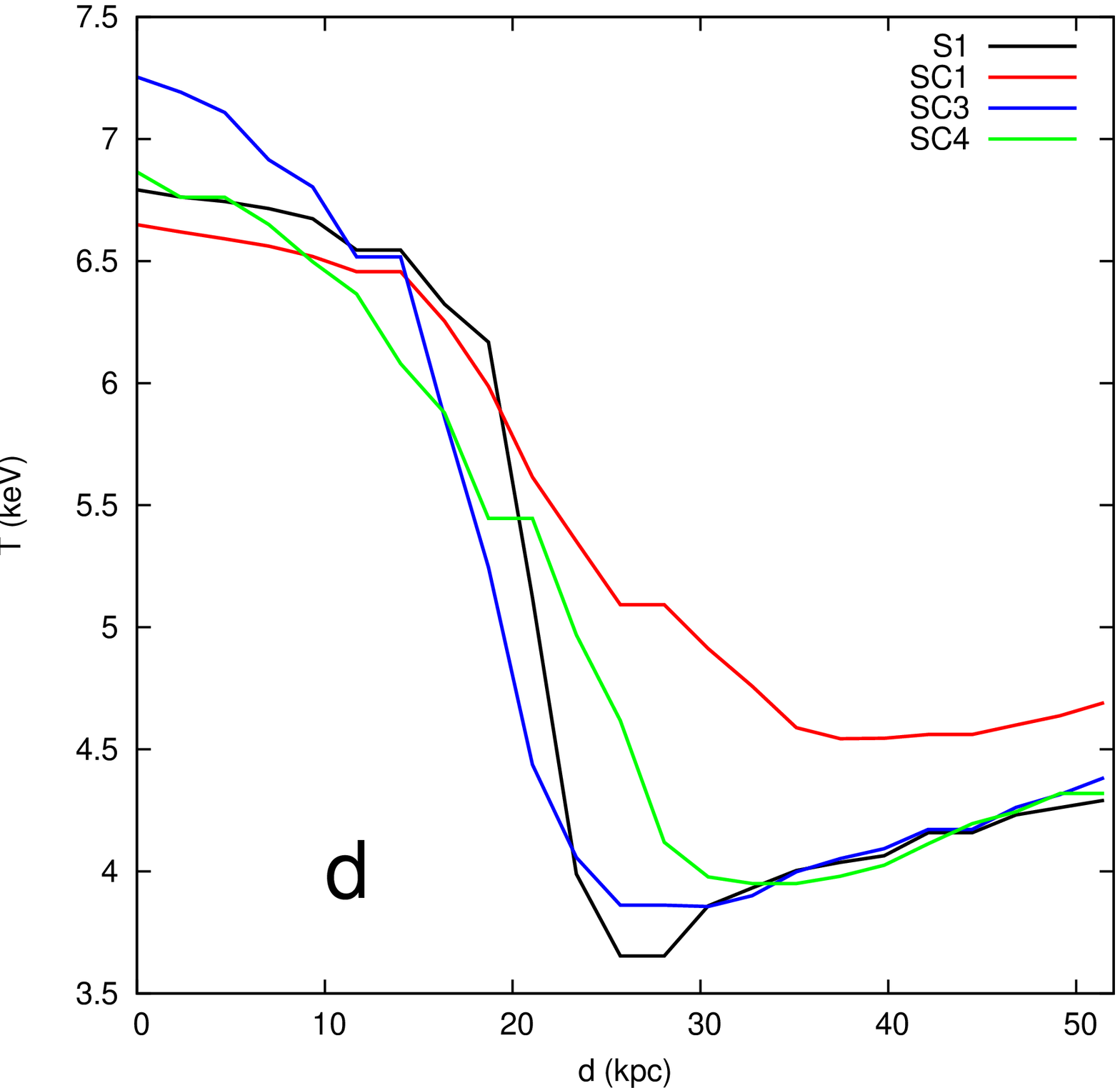}
\caption{Temperature profiles of cold fronts in simulations without
  conduction and with varying prescriptions for conduction, along the profiles marked with the
  corresponding letters in Figure \ref{fig:t3.25}. Conduction reduces
  the magnitude and increases the width of the jumps to varying
  degrees.\label{fig:temperature_jumps}}
\end{center}
\end{figure*}

Figure \ref{fig:temperature_jumps} compares
temperature gradients across cold fronts between the {\it S1}
and the {\it SC1} simulations at a late epoch, $t$ = 3.25~Gyr. Four
particular locations across the fronts are picked out for inspection
(see Figure \ref{fig:t3.25} for the profile locations, marked by the
corresponding letters). The profiles show that the temperature jumps in the conducting
simulation have all been reduced. For some of the profiles (e.g., (a)
and possibly (d)), the width of the temperature jump in the simulation with conduction {\it SC1} is not greater than the jump in the non-conducting simulation {\it S1}
(a few simulation cells, $\sim 5-10$~kpc--essentially unresolved by
the simulations), whereas for other profiles in simulation {\it SC1}, 
the temperature jumps have been smoothed out over several zone sizes
(we will explore these issues further in Section
\ref{sec:bfields}). Particularly apparent in profile (d) is that
conduction has modified the temperature distribution below the front,
while preserving the sharp jump across the front. 

\begin{figure*}
\begin{center}
\includegraphics[width=0.45\textwidth]{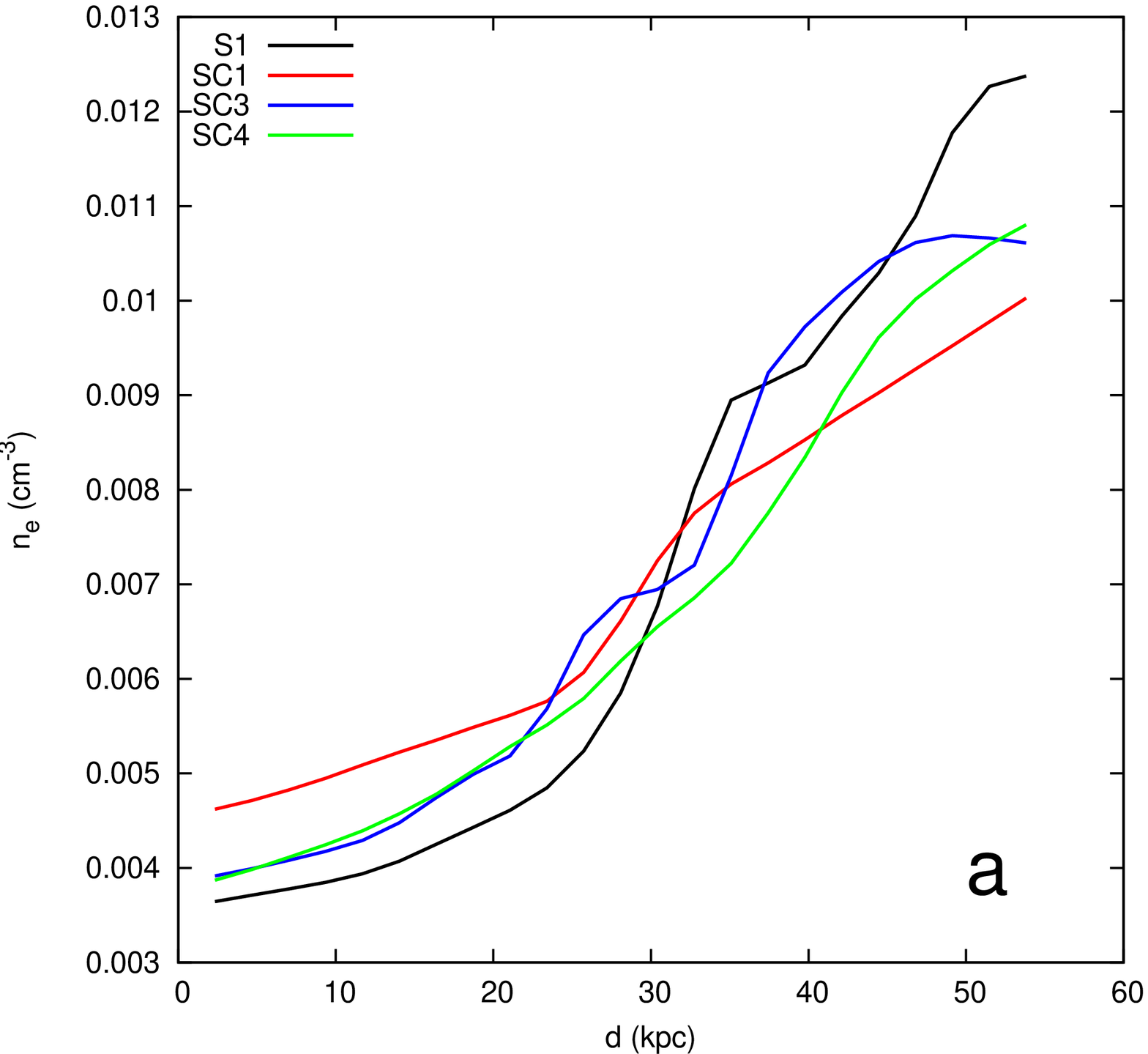}
\includegraphics[width=0.45\textwidth]{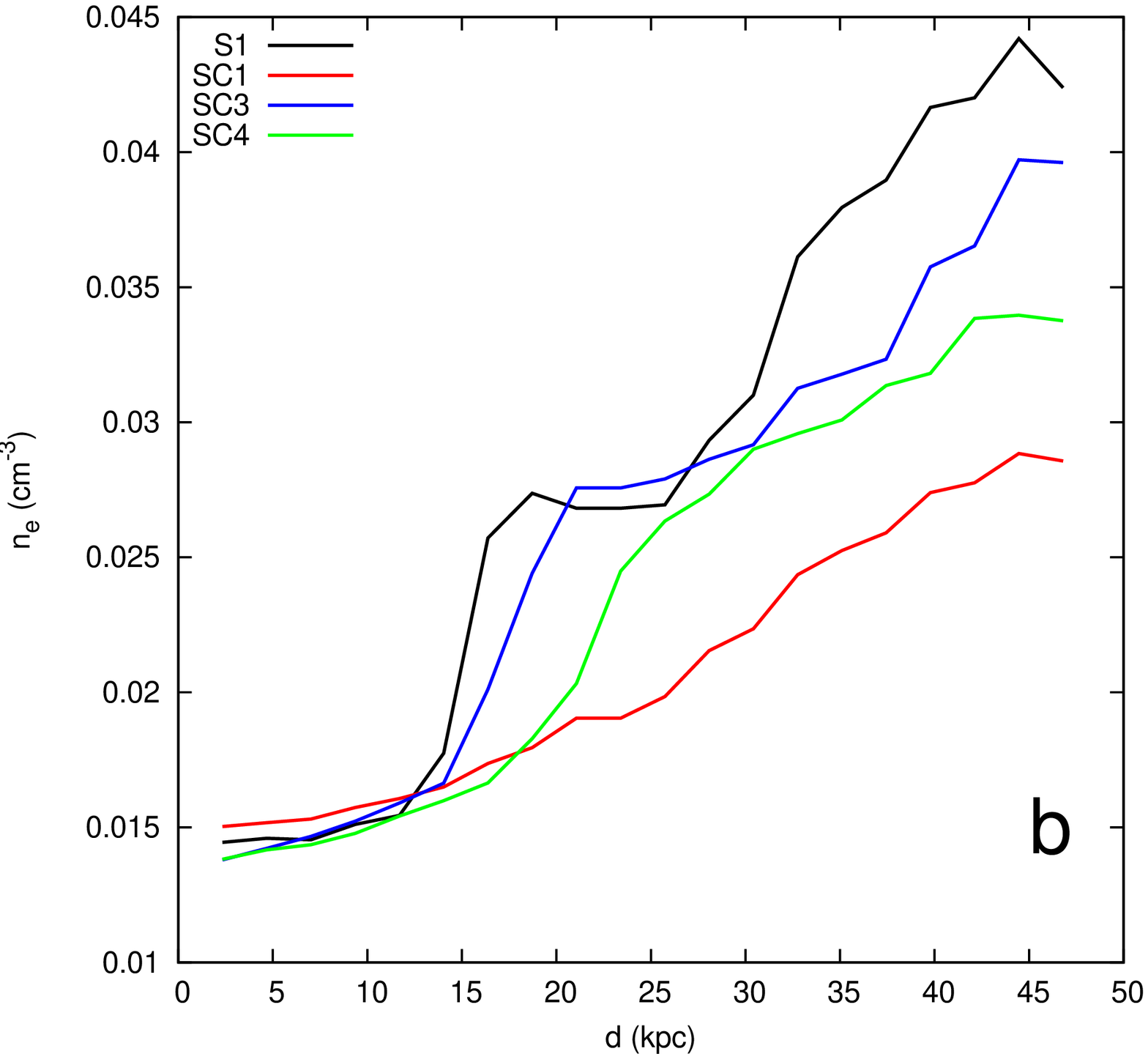}
\par
\includegraphics[width=0.45\textwidth]{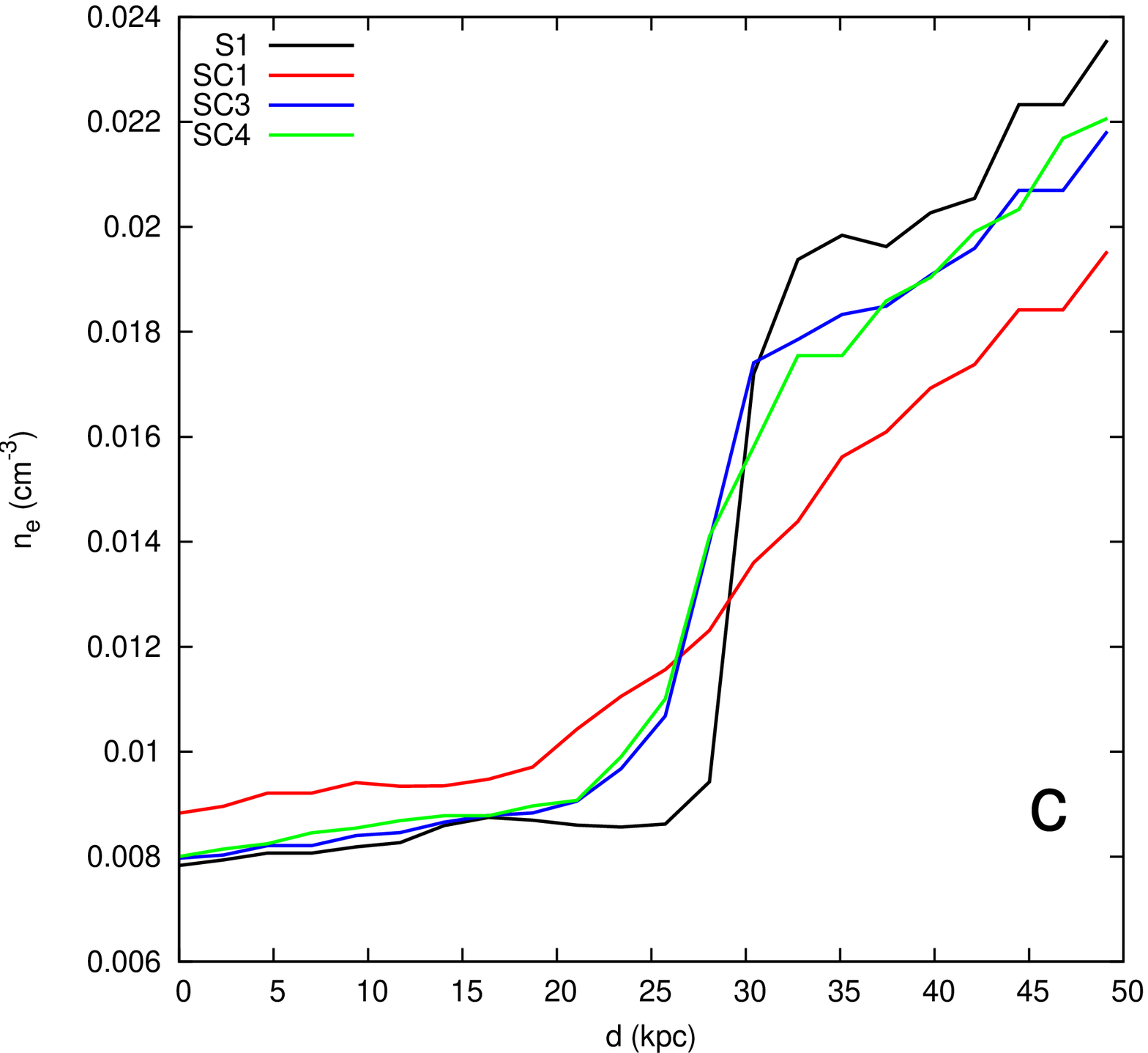}
\includegraphics[width=0.45\textwidth]{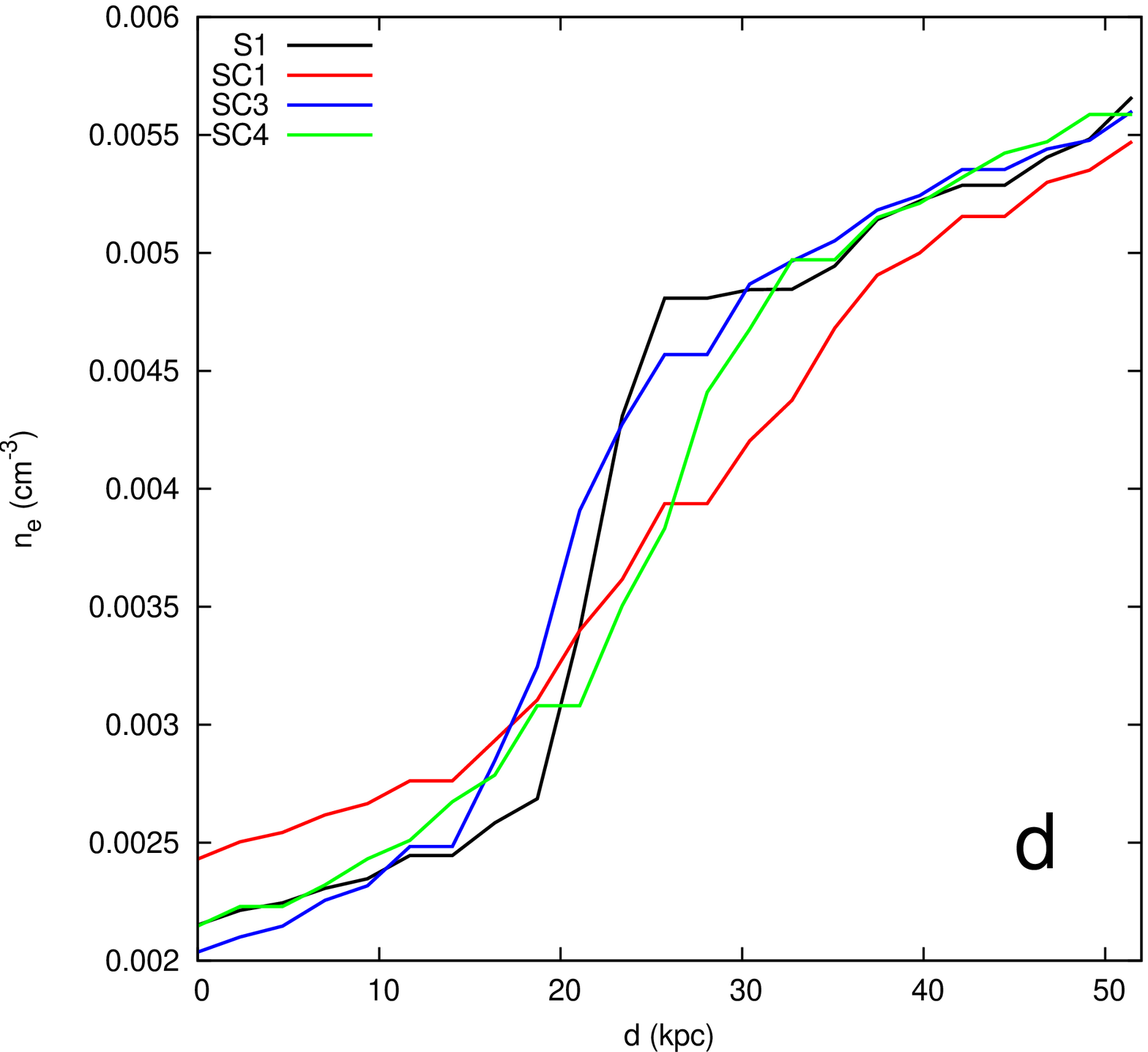}
\caption{Density profiles of cold fronts in simulations without
  conduction and with varying prescriptions for conduction, along the
  profiles marked with corresponding letters in Figure
  \ref{fig:t3.25}.\label{fig:density_jumps}}
\end{center}
\end{figure*}

Figure \ref{fig:temperature_jumps} also highlights example temperature jumps
in the {\it SC3} and {\it SC4} simulations (along the same profile locations, see Figure
\ref{fig:t3.25}). The temperature difference across the fronts are
similar between these two simulations and the {\it S1} simulation, due
to the high suppression of conduction, but the smoothing out of the gradient due to the conduction perpendicular to the field lines in the {\it SC4} simulation is noticeable, with the temperature change occurring over roughly twice the number of cells ($\sim$10~kpc) compared to the case without perpendicular conduction due to field line reconnection. 

\begin{figure*}
\begin{center}
\includegraphics[width=0.45\textwidth]{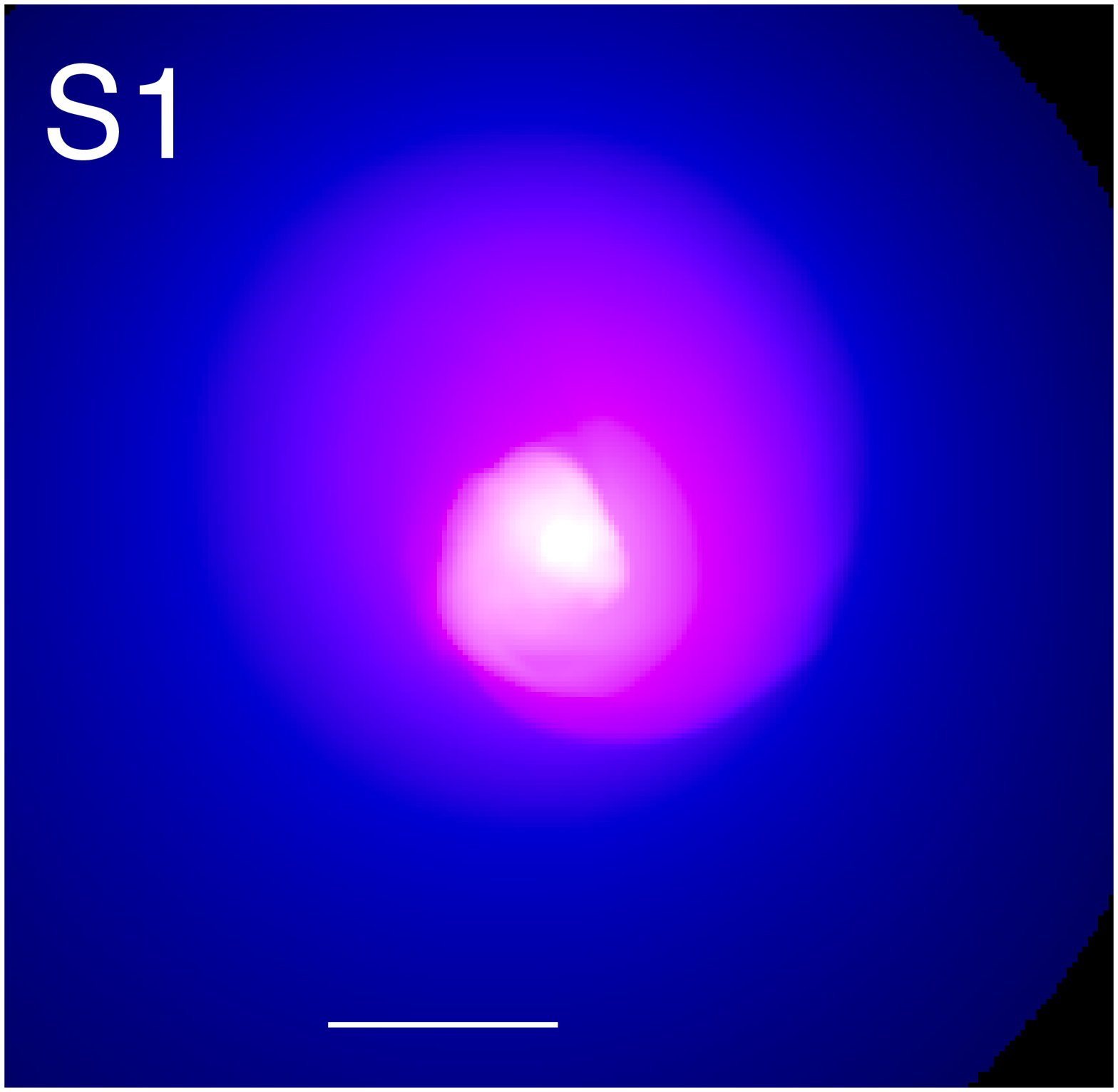}
\includegraphics[width=0.45\textwidth]{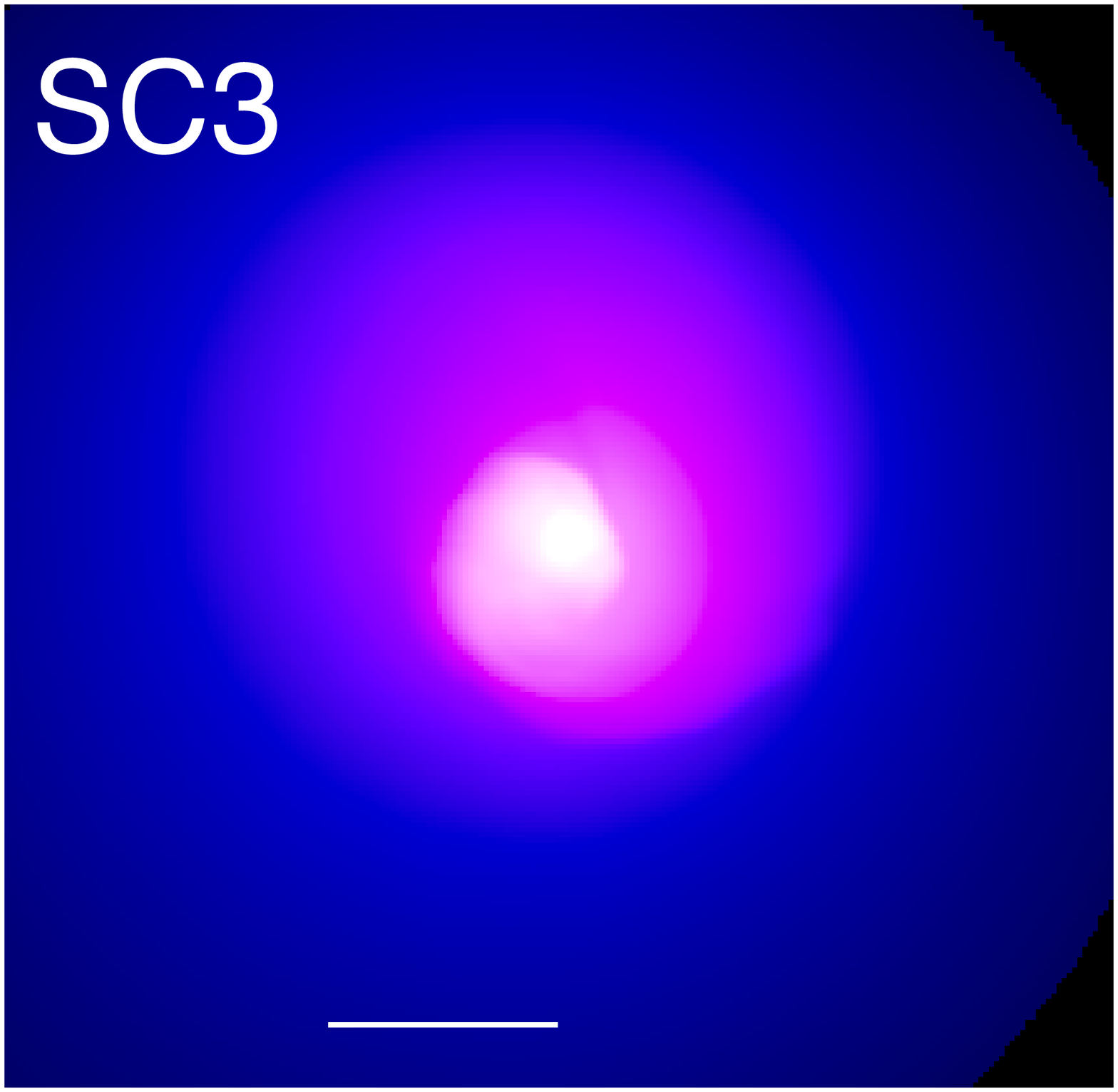}
\par
\includegraphics[width=0.45\textwidth]{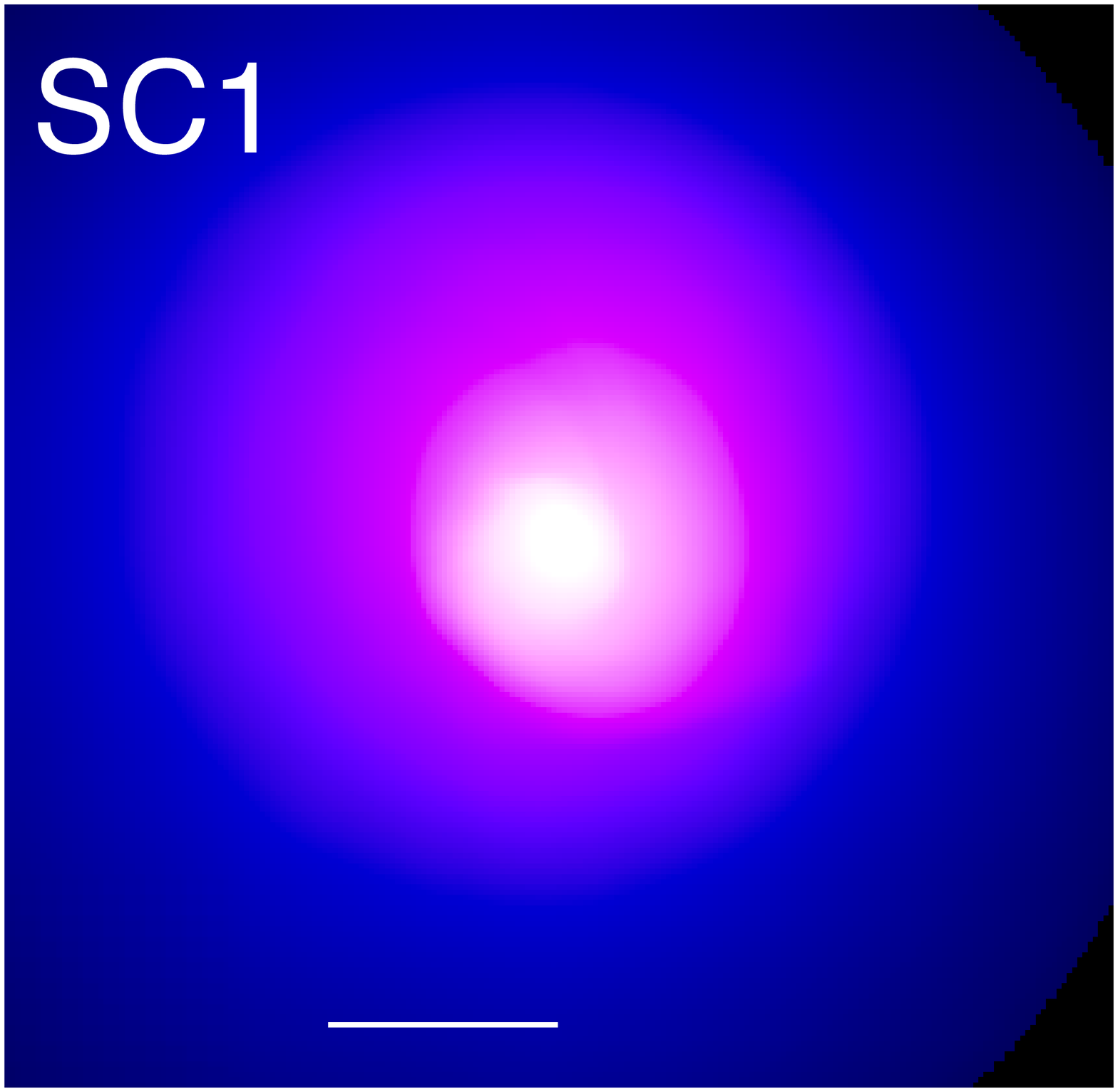}
\includegraphics[width=0.45\textwidth]{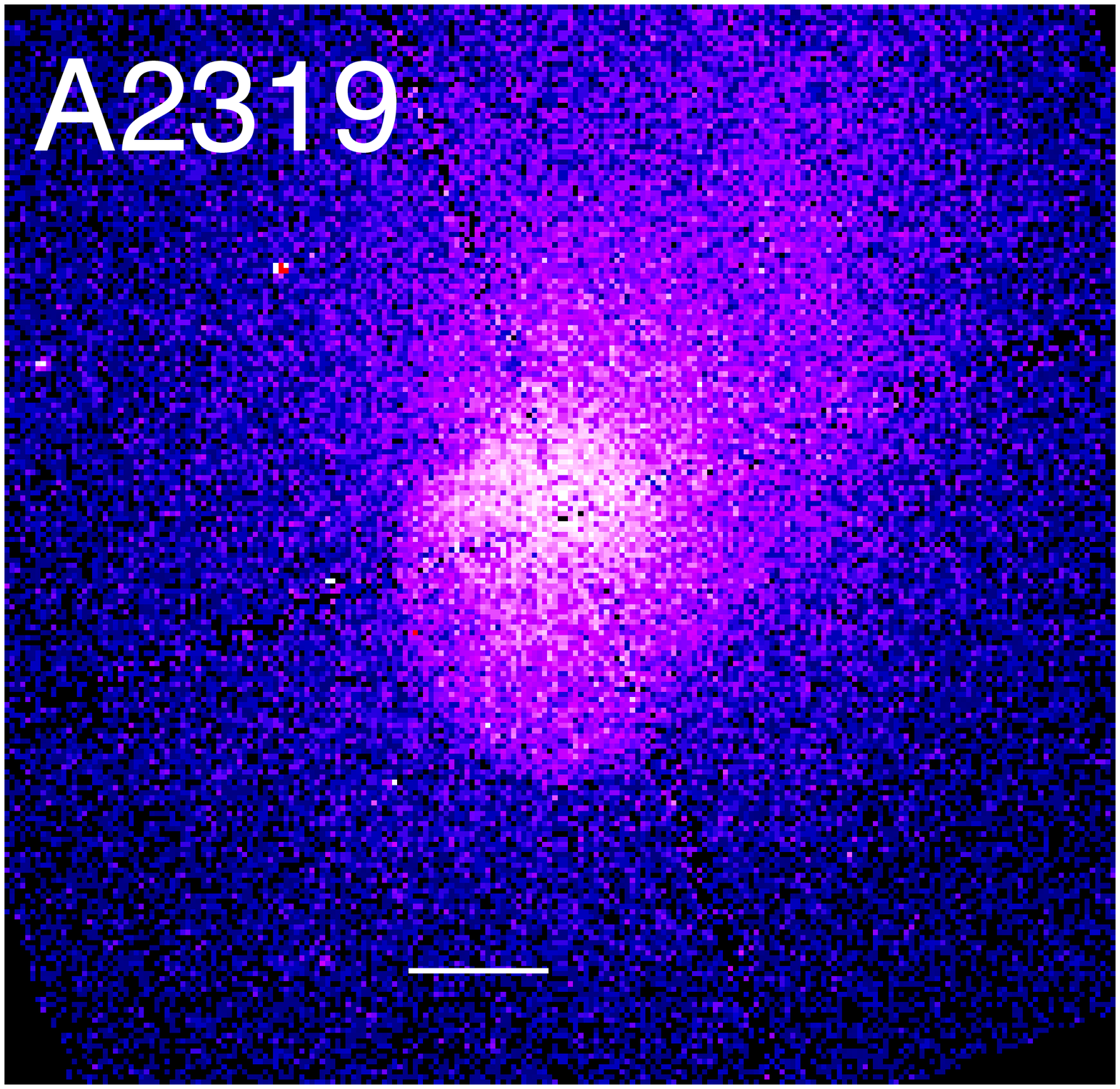}
\caption{Projected X-ray emission along the $z$-axis of the
  simulation domain for the {\it S1} (no conduction), {\it SC1}
  (Spitzer conduction), and {\it SC3} (0.1 Spitzer conduction)
  simulations at the epoch $t$ = 2.75~Gyr, with a {\it Chandra} X-ray image of A2319 included
  for comparison. White bars indicate 100~kpc distances. This
  indicates that conduction smooths out the density jumps, making them
  barely discernable as seen in X-ray images in comparison to the
  sharp jumps in emission seen in A2319.\label{fig:xray_images}}
\end{center}
\end{figure*}

The jumps in density across the cold fronts are similarly affected by
conduction. As heat flows from the hot gas above the cold fronts to
the cool gas below the front, the former contracts and the latter expands in the same proportion as
the change in temperature. Figure \ref{fig:density_jumps} shows the density
jumps across the cold fronts at the same positions in Figure
\ref{fig:t3.25}. In simulations with conduction, the density jumps are
reduced and the widths of the density jumps are increased to varying
degrees, in the same proportion as the temperature jumps except that
the sign of the gradient is reversed. In the most extreme cases, the
density jumps have been essentially erased, either in the strong
parallel conduction case {\it SC1} or the case {\it SC4} with a small
component of conduction perpendicular to the magnetic field lines. The smoothed
profiles in these simulations are in disagreement with observations of
cold fronts in clusters. 

The differences between the simulations {\it S1}, {\it SC3}, and {\it SC4} indicate that numerical reconnection across the cold front surfaces and any associated heat flux due to this reconnection is negligible. The cold fronts in the simulations {\it S1} and {\it SC3} have essentially the same temperature and density jumps, whereas adding a small perpendicular conduction smooths out the jumps over twice the number of cells. The fact that we do not see this degree of smoothing in the {\it SC3} simulation indicates that numerical reconnection is too small to provide any significant heat flux across the fronts. 

The smaller density jumps and the modification of the density profiles
below the fronts that reduce the density contrast have a
significant effect on the appearance of cold fronts as seen in X-ray
images, due to the strong dependence of the X-ray emission on the gas
density. Figure \ref{fig:xray_images} shows synthetic X-ray images for the {\it
  S1}, {\it SC1}, and {\it SC3} simulations, at the epoch $t$ = 2.75~Gyr, created by
computing the bolometric MeKaL emissivity in each FLASH cell and projecting it
along the $z$-axis (in the same manner as ZML11). Additionally, a {\it
  Chandra} X-ray image of Abell 2319 is also presented for comparison
(this cluster is slightly hotter than our model cluster, with average $T \sim
9-10$~keV). This cluster exhibits a cold front with a very sharp jump
in X-ray emission, with a width roughly the size of the {\it Chandra}
PSF \citep{oha04,gov04}. 

The characteristic spiral shape of the cold fronts is apparent in each
simulated image. However, the sharp edges in emission across the front surfaces
are essentially absent from the {\it SC1} case, whereas in the {\it SC3} and {\it S1} simulations, the sharp
jumps are clearly seen. Therefore, for our simulated hot cluster,
unsuppressed Spitzer conduction appears to produce cold fronts that
are far less prominent than those in observed hot clusters, and it is
questionable whether or not they may be even called ``cold fronts'' at
all. We reserve a more quantitative comparison with observed clusters of different temperatures and plasma $\beta$ values and the possible constraints on conductivity for a future paper. 

It also is apparent from these figures that conduction may have an effect on the
smoothness of the front surfaces. For example, the cold fronts in
simulations {\it SC1} and {\it SC2} exhibit fewer small-scale features
along the front surfaces than their counterparts without conduction,
{\it S1} and {\it S2}. The widening of the cold front interfaces (seen
clearly in Figure \ref{fig:density_jumps}) will cause them to be less
susceptible to the Kelvin-Helmholtz instability \citep[see,
e.g.,][]{chu04}.

\subsection{The Geometry of the Magnetic Field Lines\label{sec:bfields}}

\begin{figure*}
\includegraphics[width=0.993\textwidth]{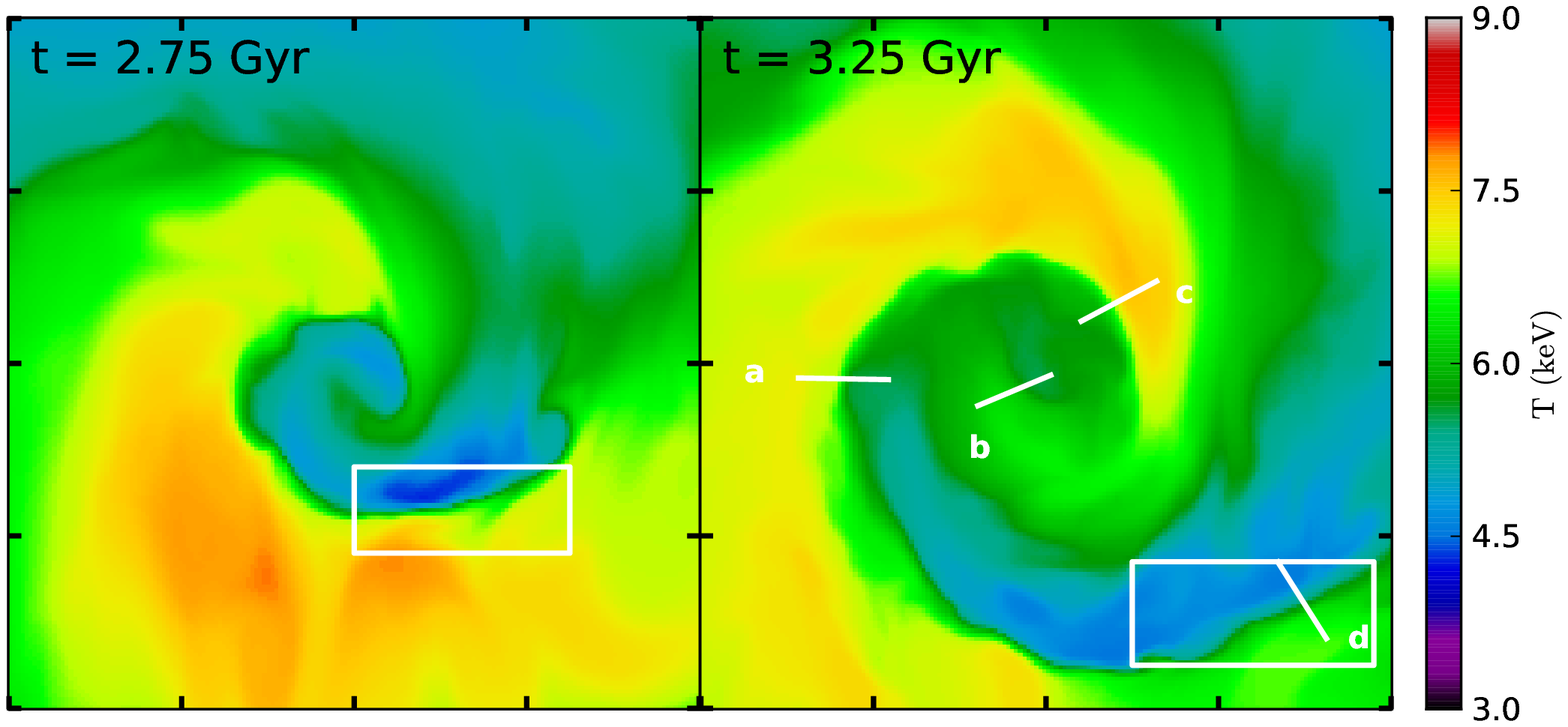}
\par
\includegraphics[width=\textwidth]{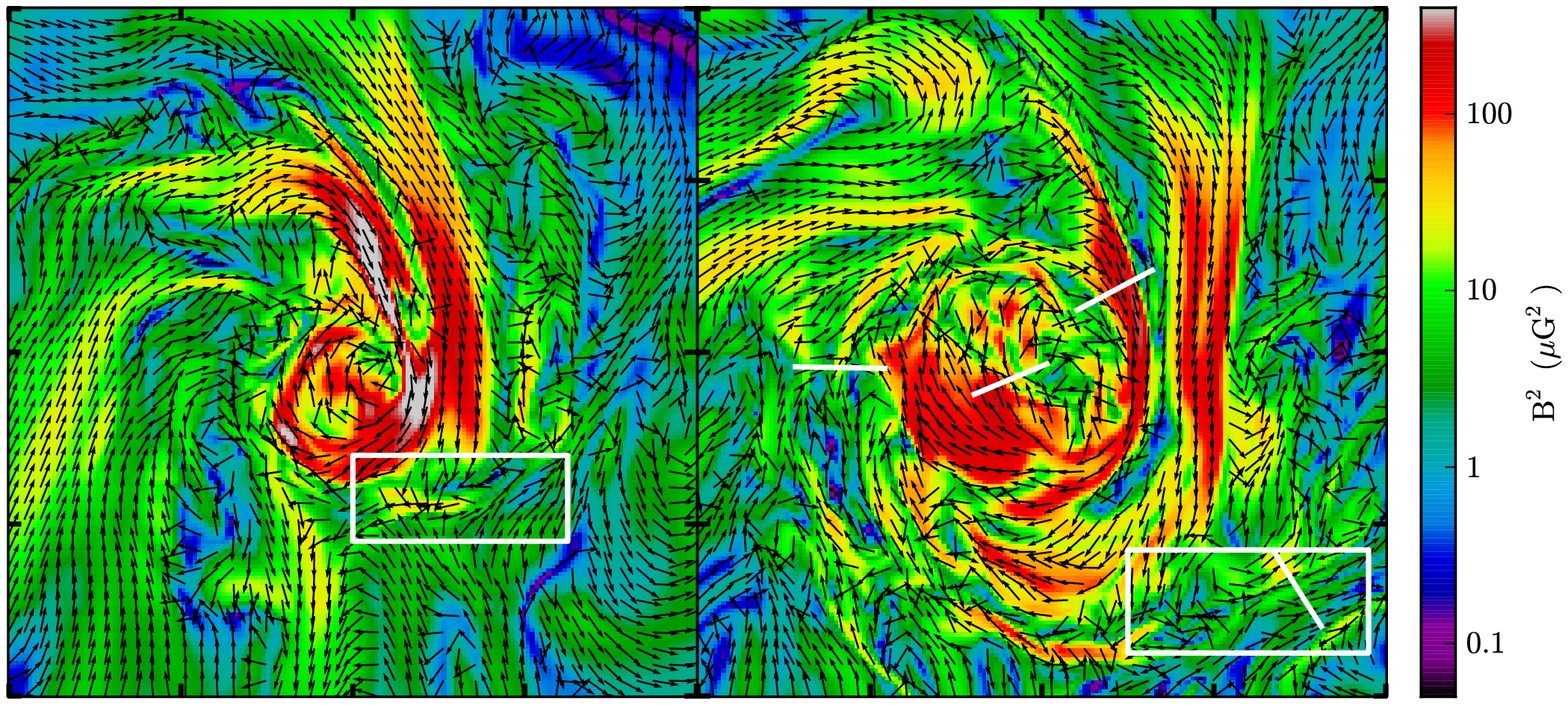}
\caption{Slices through the $z$ = 0 coordinate plane of the
  temperature and the magnetic field energy in the vicinity of four
  representative profiles with the 2D vector field
  $\hat{\textbf{\emph{b}}} = (b_x, b_y)$ overlaid for the simulation
  {\it SC1} (Spitzer conduction along field lines) at the epochs $t$ =
  2.75~Gyr and 3.25~Gyr. Each panel is 400~kpc on a side. Tick marks
  indicate 100~kpc distances. The field vectors are
  located at the positions of their tails. Top panels: Temperature in keV. Bottom panels:
  The magnetic field energy $B^2$ in $\mu{G}^2$. Red, elongated regions in the bottom panels indicate magnetic draping layers. The white rectangles indicate the regions
  without strong draping layers described in the text. The white lines
  indicate the profile locations from Figure \ref{fig:t3.25}.\label{fig:vector_figs}}
\end{figure*}

\begin{figure*}
\begin{center}
\includegraphics*[width=0.468\textwidth]{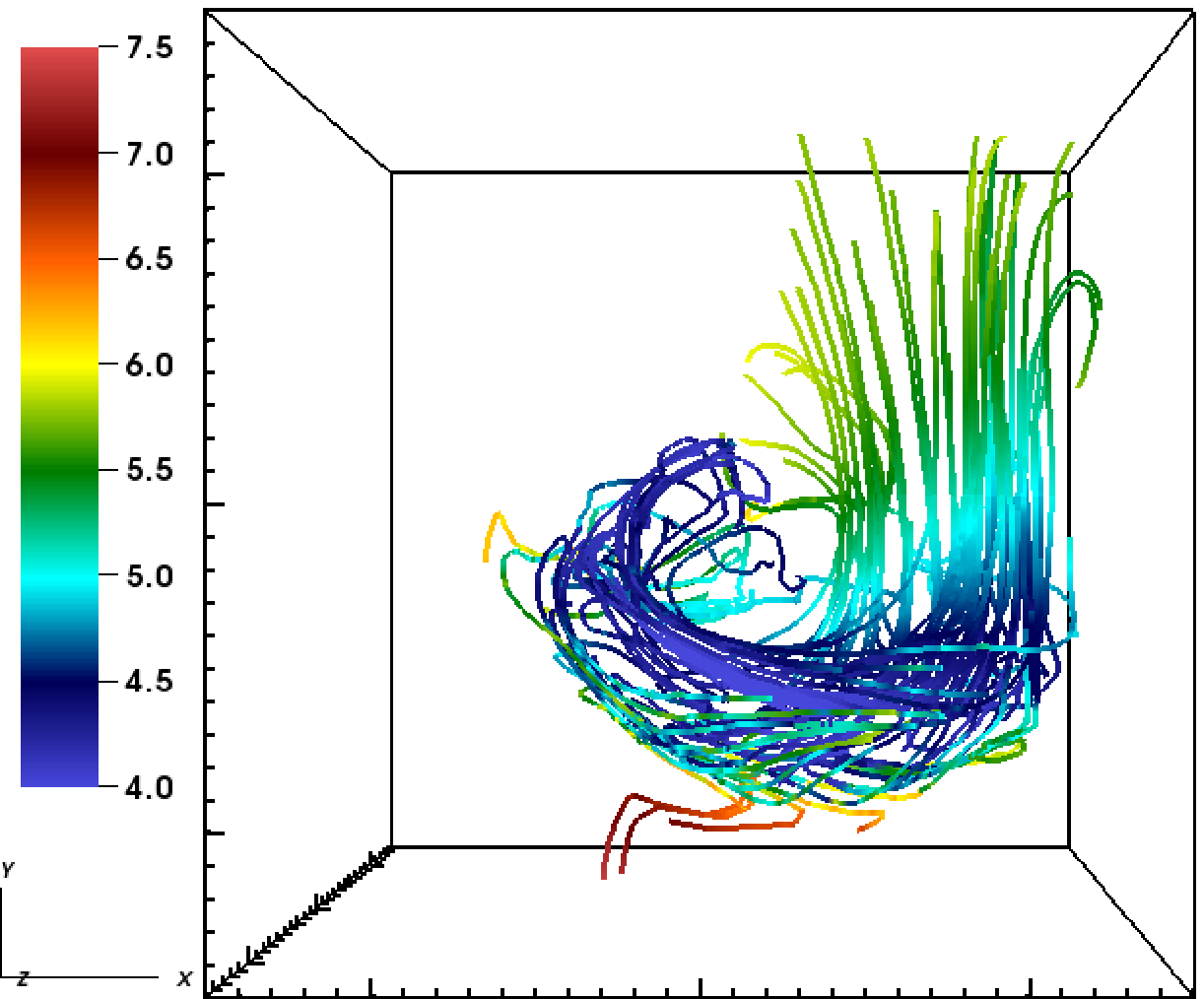}
\enspace
\includegraphics*[width=0.5\textwidth]{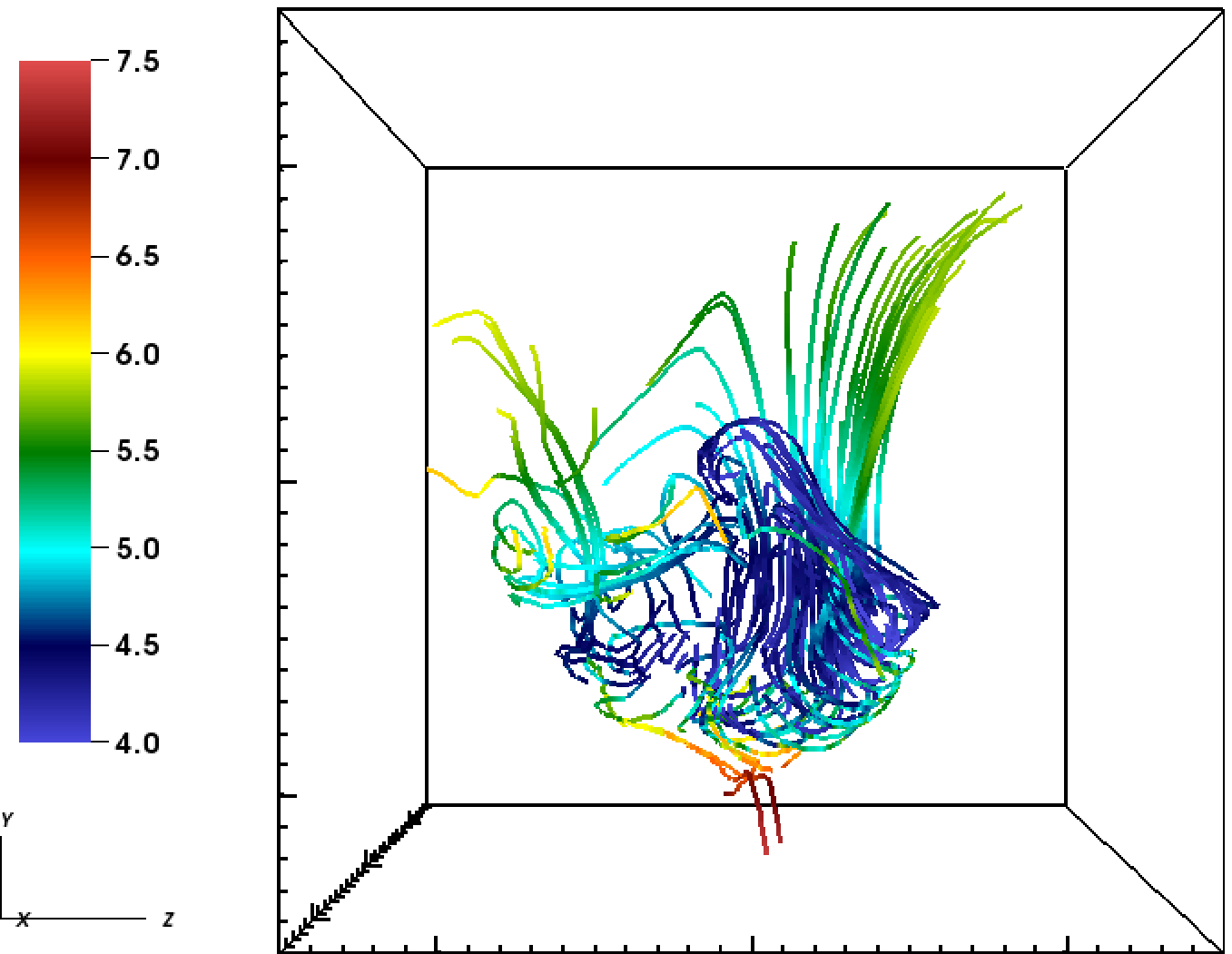}
\caption{Magnetic field lines in the {\it SC1} (Spitzer conduction)
  simulation beginning at points inside the cold fronts with $T <$ 4.5~keV and continuing up to a length of
  100~kpc at the epoch $t$ = 2.5~Gyr. Left panel: View along the $z$-axis. Right panel: View
  along the $x$-axis. Color along the lines indicates temperature in
  keV. The box size is 150~kpc, and the major tick
  marks indicate 50~kpc distances. This indicates that field lines connect
  regions of cold gas under the cold front surfaces to hot gas
  outside them.\label{fig:fieldlines_t2.5}}
\end{center}
\end{figure*}

\begin{figure*}
\begin{center}
\includegraphics*[width=0.468\textwidth]{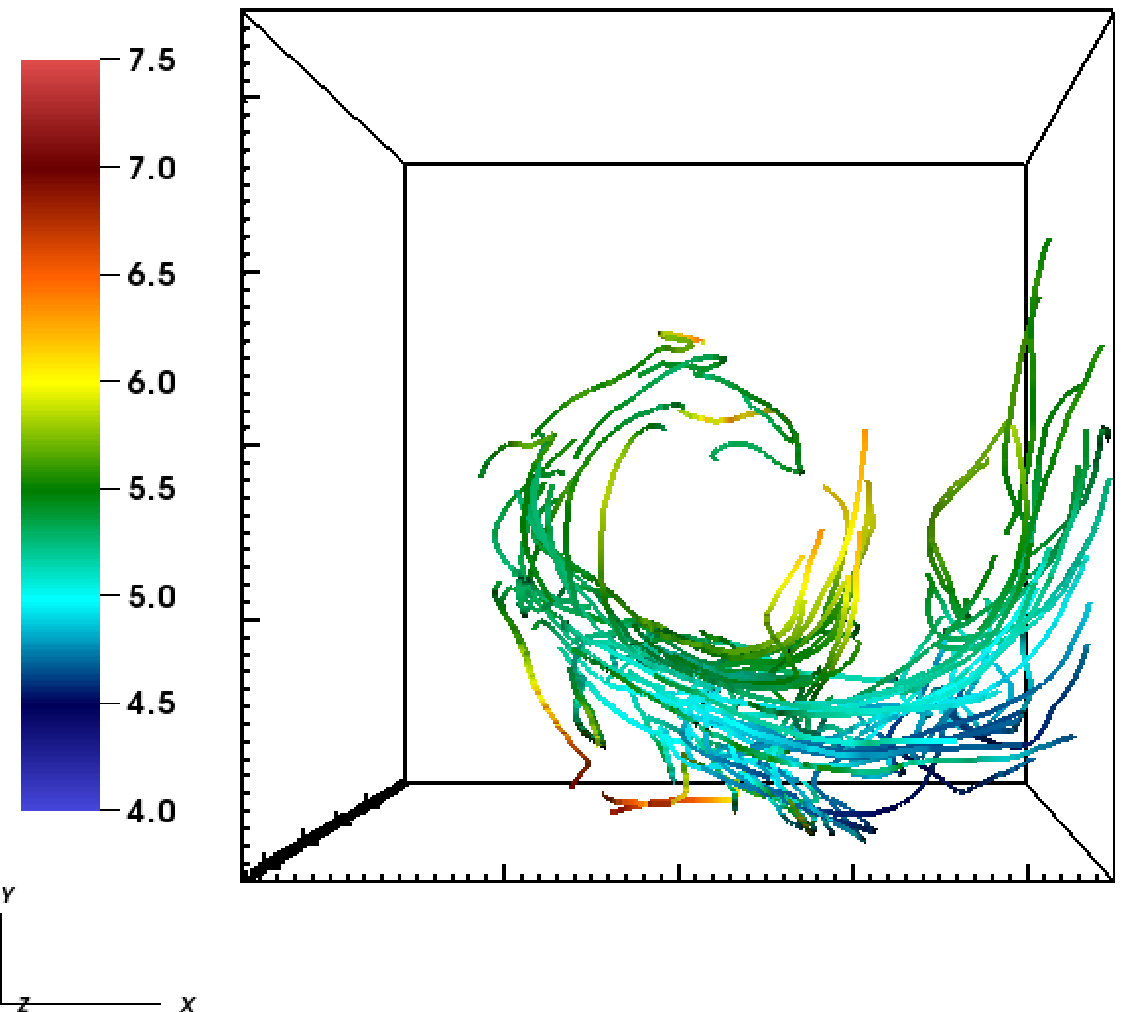}
\enspace
\includegraphics*[width=0.5\textwidth]{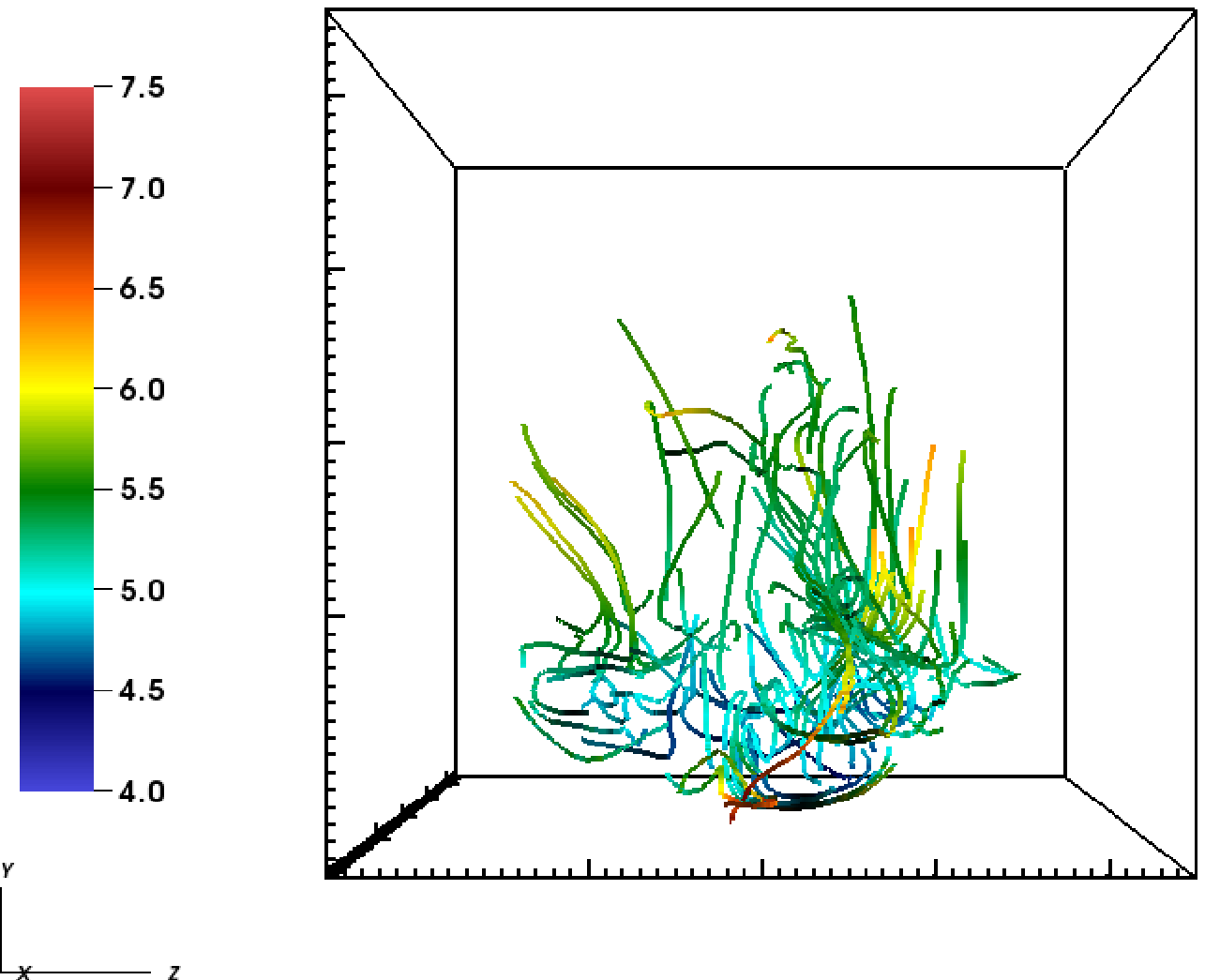}
\caption{Magnetic field lines in the {\it SC1} simulation beginning at points inside the cold
  fronts with $T <$ 5.5~keV and continuing up to a length of
  100~kpc at the epoch $t$ = 3.0~Gyr. Left panel: View along the $z$-axis. Right panel: View
  along the $x$-axis. Color along the lines indicates temperature in
  keV. The box size is 250~kpc, and the major tick
  marks indicate 50~kpc distances.  This indicates that field lines connect
  regions of cold gas under the cold front surfaces to hot gas
  outside them.\label{fig:fieldlines_t3.0}}
\end{center}
\end{figure*}

The simulations from ZML11 showed magnetic field is amplified and the
lines oriented parallel to the cold front surfaces, which
would prevent conduction across the cold front surfaces from
the hotter gas just above the fronts. The fact that there are examples
of discernable temperature jumps over the same spatial range (a few simulation
cells) even in the {\it SC1} run indicates that conduction is still strongly suppressed directly across the front
surfaces. What accounts for the significant changes in density and
temperature across the fronts that still occur?

\begin{figure}
\begin{center}
\plotone{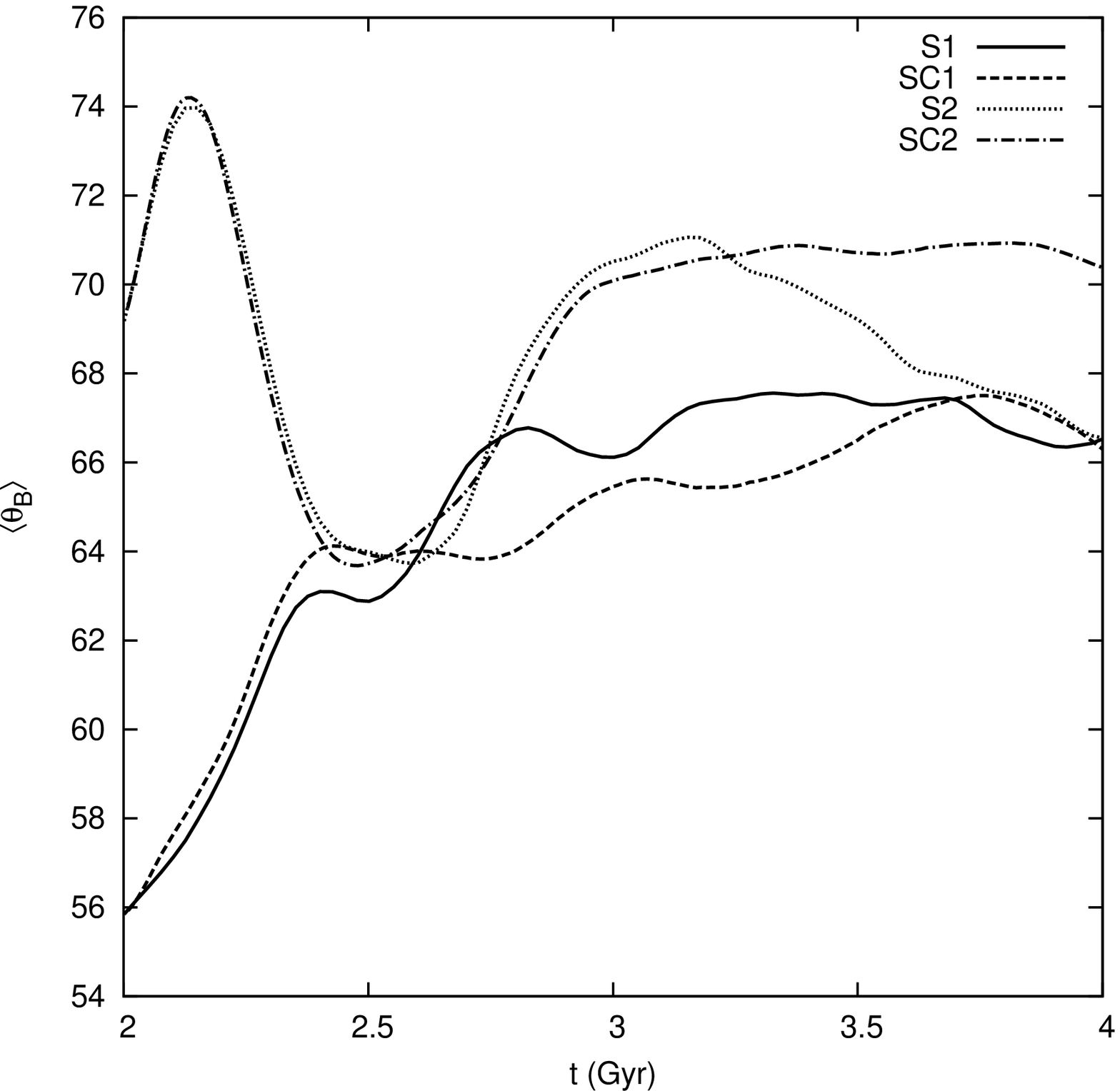}
\caption{The evolution of the volume-averaged (within a sphere of
  radius $R$ = 100~kpc) angle between the
  magnetic field unit vector and the radial unit vector $\langle\theta_B\rangle = \cos^{-1}\langle|\hat{\textbf{\emph{b}}}\cdot\hat{\textbf{\emph{{r}}}}|\rangle$ over time for
  simulations with and without conduction.\label{fig:mag_angle}}
\end{center}
\end{figure}

The cool gas below the cold fronts is only completely immune to the effects of heat
conduction if there are no field lines leading from hotter regions of
the cluster atmosphere into the cold regions of the fronts from other
directions. The fronts seen in X-ray observations do not completely enclose the cool
core. In principle, there could be magnetic field lines
that reach the regions below the fronts from hotter regions on the
other side of the cluster. More importantly, simulations of gas sloshing
show that the resulting spiral structure has an alternating pattern of
hot and cold flows, with hot flows developing below cold fronts as
they expand to higher radii. \citet{kes11} showed that these latter
flows are of hot gas being drawn into the core region form higher
radii. These regions can be magnetically connected to the cool gas
immediately below the fronts. Additionally, though the cold fronts that form in simulations
are extended in directions perpendicular to the sloshing plane, the
characteristic size of the fronts in this direction (along the line of
sight where the observer sees the spiral structure) is smaller than in
the plane. This exposes the cold regions of the fronts to magnetic
field lines extending into hotter regions along directions perpendicular to the sloshing plane. 

The preceding arguments lead us to expect that some field lines that pass through cold gas below the front
surfaces may be connected to hotter regions over relatively short
distances. Figures \ref{fig:fieldlines_t2.5} and
\ref{fig:fieldlines_t3.0} show magnetic field lines beginning at
several points in the cold region below the front surface and followed for a
distance of 100~kpc. A large number of field lines are tangentially
oriented to the front surfaces, as can be easily seen in both
projections, shielding this cold gas from the hotter gas directly
above the front. However, the figures show that there are still a number of field lines that
extend into hotter regions of the cluster core along
other directions. At the epochs of $t$ = 2.5~Gyr and $t$ = 3.0~Gyr,
the coldest temperatures below the front surfaces are approximately
$T \sim$4.5-5~keV, and these regions are connected to regions with
temperatures of $T \sim$6-7.5~keV. In the right panel of Figure
\ref{fig:fieldlines_t2.5}, field lines stretched along the $y$ and
$z$ directions connect cold gas underneath the front surfaces to
hotter gas. In Figure \ref{fig:fieldlines_t3.0}, lines connecting cool gas with
hot gas that was brought in by sloshing are evident near the center of the
left panel of the figure. 

Finally, the simulations show that there is not always a perfect alignment of strong magnetic field layers with cold fronts. Figure \ref{fig:vector_figs} shows slices through the center of the simulation domain in the $x-y$ plane (coincident with the plane of the
sloshing motions) of the gas temperature and the magnetic field energy for the epochs
$t$ = 2.75~Gyr and $t$ = 3.25~Gyr with the magnetic field vector components in the $x$ and
$y$-directions overlaid. Extended regions of strong magnetic field indicate
places where the magnetic field strength has been increased by shear
amplification. The locations of the profiles from Figure
\ref{fig:t3.25} are also marked. While some locations along the cold fronts
exhibit a strong field layer parallel to the front, this is
not strictly true everywhere along the fronts. In particular, the $t$
= 2.75~Gyr panel shows that the trailing edge of the cold spiral in
the southwestern direction (indicated roughly by the white rectangles in Figure
\ref{fig:vector_figs}) is not draped by a high-$B$ layer. It
can be seen that the magnetic field vectors along this portion of the
spiral front are not strictly aligned parallel to the front
surface. Regions of ordered, strong field are interspersed with
regions of weak, tangled field as shown by the vectors at these
locations. At a somewhat later time, $t$ = 3.25~Gyr, this location still
lacks a strongly magnetized parallel layer. This also happens to be the
location of profile (d). That particular temperature jump was shown to be
smoothed out significantly in Figure \ref{fig:temperature_jumps}. Though at the epoch $t$ = 3.25~Gyr there does appear to be a small, parallel, highly magnetized layer at that particular front location, this layer was not present previously and the temperature
jump likely has been smoothed out before the formation of this layer. 

Cold fronts that are not draped by strong magnetic fields
will be more susceptible to the K-H instability, which will distort the front
surface and lead to gas mixing between hot and cold
phases. This is because the orientation of the draping fields tends to
coincide with the direction of the shearing motions. Perturbations of the field lines perpendicular to the front
surfaces at these locations will lead to smoothing out of the fronts
as outlined in \citet{lec12}, though the extent of this smoothing will
be limited by the HBI, which will reorient magnetic field lines
parallel to the front surface. \citet{lec12} estimated that the scale height of
the temperature gradient for such a front would smooth to a height of
$h_T \sim \chi^{2/3}/g^{1/3}$, where $\chi$ is the Spitzer thermal
diffusivity and $g$ is the gravitational acceleration. Taking example
values from our simulation for a cold front at $r \approx$120~kpc, we
find $\chi \sim 5 \times 10^{30}$~cm$^2$~s$^{-1}$ and $g \sim 3 \times
10^{-8}$~cm~s$^{-2}$. This yields an approximate scale height of the
temperature gradient across the front of $h_T \sim 30$~kpc, which is
similar to the widths of some of the temperature gradients across the fronts
in Figure \ref{fig:temperature_jumps}. 

In simulations of thermal instabilities in magnetized galaxy clusters, a helpful
quantity to examine has been the average angle between the magnetic field
and radial unit vectors, given by
\begin{equation}
\langle\theta_B\rangle = \cos^{-1}\langle|\hat{\textbf{\emph{b}}}\cdot\hat{\textbf{\emph{{r}}}}|\rangle
\end{equation}
Isotropically oriented tangled field lines will have an average
$\langle|\hat{\textbf{\emph{b}}}\cdot\hat{\textbf{\emph{{r}}}}|\rangle
= 0.5$ and hence the corresponding angle is $\langle\theta_B\rangle \approx
60\degree$. The HBI drives field lines in a more azimuthal direction
($\theta_B \sim 70-80\degree$) in the presence of a positive radial
temperature gradient (as in a cool core). Sloshing motions in general preserve the overall
positive temperature gradient of the cool core, implying that the HBI
should still be active. However, they will also drag the field lines
around into a more azimuthally oriented configuration as they are
draped around the cold fronts. 

Figure \ref{fig:mag_angle} shows the evolution of the volume-averaged angle of
the magnetic field with respect to the radial direction within a
radius of 100~kpc over time for the {\it S1}, {\it S2}, {\it SC1},
and {\it SC2} simulations. The simulations with initially tangled
and azimuthal fields begin with $\langle\theta_B\rangle \approx 56\degree$ and
$\langle\theta_B\rangle \approx 70\degree$, respectively. We find that
largely the evolution of this ``magnetic angle'' is driven by the
sloshing motions themselves, since the evolution of $\theta_B$ is
essentially identical between the simulations with and without
conduction that have the same initial magnetic field setup. An
exception to this general rule are the simulations with initially
azimuthal magnetic field lines, {\it S2} and {\it SC2}. When
conduction is included, the final $\theta_B$ indicates a more
azimuthal direction than in the case where conduction is absent. This is possibly due to the
fact that in this case the core is more magnetically isolated due to
the azimuthally oriented field lines, and the positive temperature
gradient is better preserved in the core than in the tangled field
line case. In such a scenario,  the HBI would drive
the field lines in a more azimuthal direction once the conduction is
turned on. 

\subsection{The Evolution of the Entropy of the Cluster Core\label{sec:entropy_evolution}}

\begin{figure}
\begin{center}
\plotone{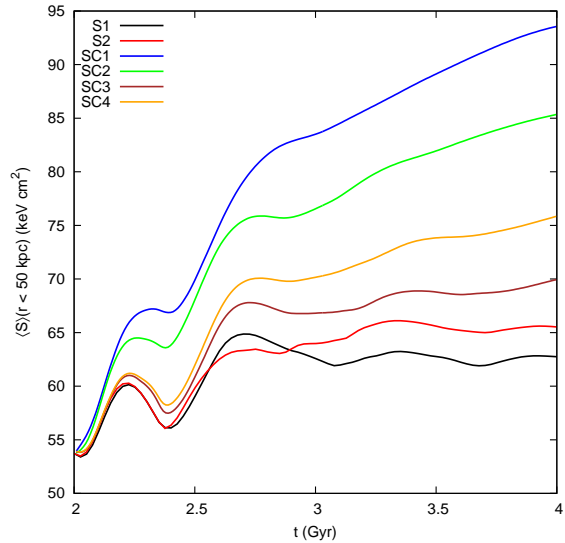}
\caption{Evolution of the average entropy within 50~kpc of the cluster center
  for simulations with and without anisotropic thermal conduction.\label{fig:echange_conduction}}
\end{center}
\end{figure}

Given that our simulations show that the magnetic field geometry of
the sloshing cool core still permits heat conduction to the cluster core, it is
important to determine what effect this has on the thermal state of
the core. The evolution of average entropy within the inner
50~kpc of the cluster center over time is shown in Figure
\ref{fig:echange_conduction}. In the simulations without conduction
({\it S1} and {\it S2}), the average entropy of the cluster core
increases due to sloshing bringing hot gas from the cluster outskirts
with the cool gas of the core into contact, which then mix to a
certain degree. However, the magnetic field suppresses such mixing
(ZML11), and the increase in entropy levels off roughly 0.5~Gyr after
the simulation begins. ZML11 showed that the
difference in initial field geometry has little effect on the entropy
increase in this case.

In contrast, in the simulations with anisotropic Spitzer conduction
({\it SC1} and {\it SC2}), heat conducts along
the field lines to the cluster center, and the entropy increase of the
core continues unabated for the duration of the simulation. As
expected, the effect is stronger in the case with initially random field lines than in the
case with initially azimuthal field lines. The average specific entropy increase in
these two simulations is roughly $\Delta{S} \sim 20-30$~keV~cm$^2$
higher than in the adiabatic simulations.

A similar situation exists for the simulations with weaker conduction
coefficients. The simulation with conductivity 1/10 of the Spitzer
value strictly along the field lines ({\it SC3}) has a correspondingly smaller increase in
entropy over the {\it SC1} simulation, with $\Delta{S} \sim 5$~keV~cm$^2$, close to the
entropy increase due to sloshing alone, but the entropy is still
increasing at the end of the simulation. The inclusion of even the
small, 1/100 Spitzer conduction perpendicular to the field lines ({\it
  SC4}) results in a significantly higher entropy increase over
simulation {\it S1} than the {\it SC3} case, with $\Delta{S} \sim
10$~keV~cm$^2$--this is because in the sloshing core, temperature
gradients across the magnetic field lines are much higher than along
the lines. The result of all of the simulations with some form of
anisotropic conduction is that heat can be conducted efficiently to
the core, even in the presence of azimuthal cold fronts. 

\subsection{Simulations with Conduction and Radiative Cooling\label{sec:cooling}}

\begin{figure*}
\begin{center}
\includegraphics[width=0.8\textwidth]{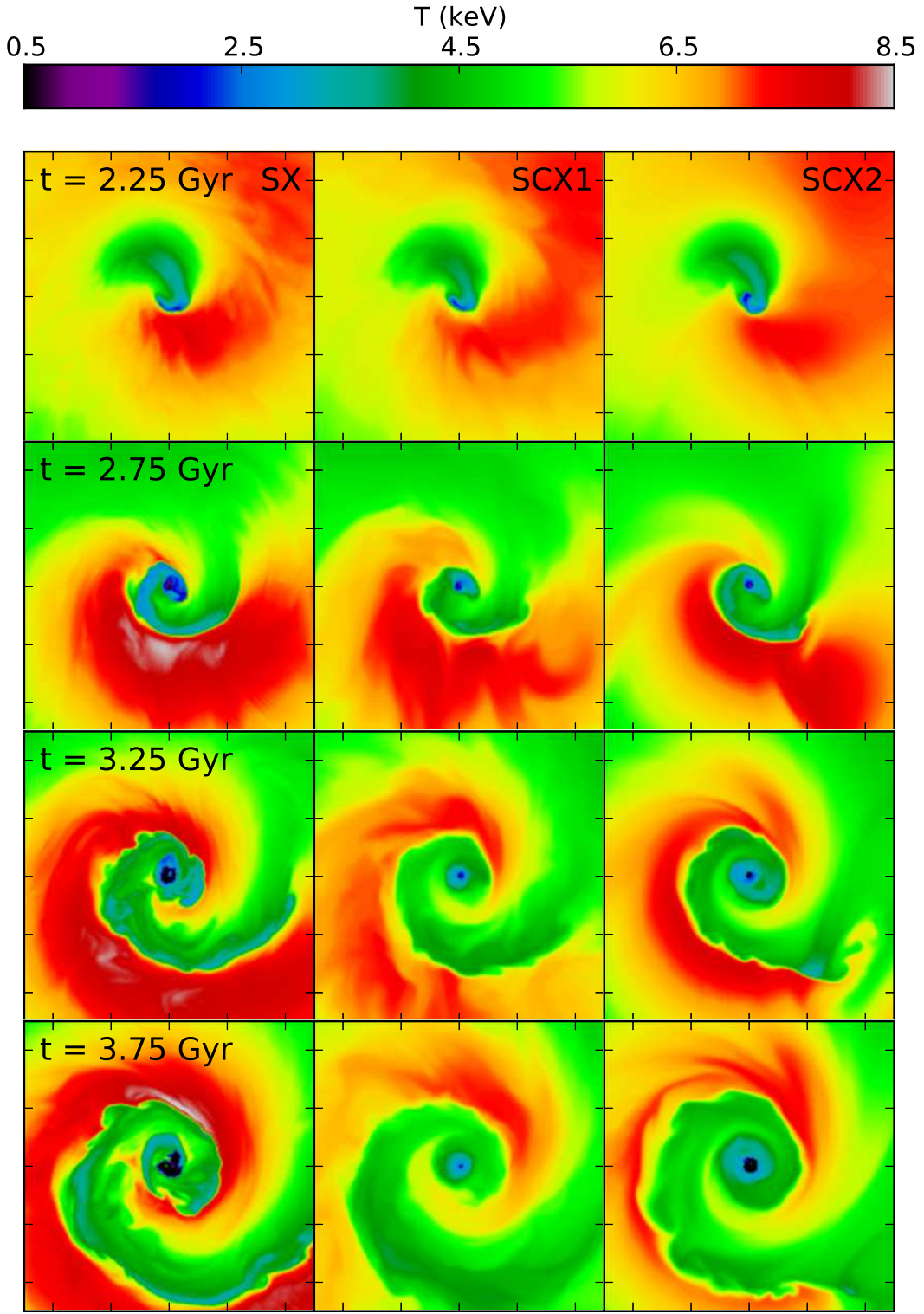}
\caption{Slices through the center of the domain in the $x-y$ plane of the temperature for the sloshing simulations
  with cooling for the epoch $t$ = 2.25, 2.75, 3.25, and 3.75~Gyr. Each panel is 500~kpc on
  a side. The colorscale is temperature in keV. Major tick marks
  indicate 100~kpc distances.\label{fig:cooling}}
\end{center}
\end{figure*}

We now include radiative cooling in our simulations. The runs with
radiative cooling {\it SCX1} and {\it SCX2} correspond to the
non-cooling runs {\it SC1} and {\it SC2}. The cooling is ``turned on'' at the same moment as
the anisotropic conduction, at the epoch $t$ = 2~Gyr. As control cases to isolate the effects of sloshing
and conduction, we also ran the following simulations. {\it RX} is an initially relaxed cluster left in
isolation without heat conduction. Due to the absence of any heating mechanism, it experiences
a cooling catastrophe in the core within a few hundred Myr. {\it RCX}
is a similar run with an isolated cluster, except it also includes
Spitzer conduction along the field lines. Thermal conduction partially offsets the cooling of the
core in this run, but the HBI quickly reorients the magnetic field
lines in the core in the azimuthal direction and a cooling
catastrophe still occurs. The last run, {\it SX}, is identical to
simulation {\it S1} (without conduction and with
sloshing), but with radiative cooling included.

In ZMJ10, we showed that radiative cooling may be slowed by sloshing itself, due to
the mixing of hot gas from the outskirts with the cold gas for the
core. However, the inclusion of even weak
magnetic fields reduces this mixing and inhibits the consequent
heating of the core, depending on the strength of the magnetic
field (ZML11). From our results in ZMJ10, we expect that regardless of
the strength of the magnetic field, the heat delivered to the core
from sloshing alone will not be sufficient to prevent a cooling
catastrophe at the innermost regions of the cool core. The left panels
of Figure \ref{fig:cooling} shows slices of temperature through the
center of the cluster for selected epochs for the {\it SX}
simulation. In this run, the core temperature drops below 1~keV within
a radius of $r \sim 20$~kpc. Additionally, the core density peaks at
high values of $n_e \sim 1-5$~cm$^{-3}$, in conflict with observations
of real clusters. Therefore, the goal of the runs with conduction and
cooling is to determine if sloshing with conduction together is able
to prevent such a cooling catastrophe. 

The center and right panels of Figure \ref{fig:cooling} shows slices of temperature through the
center of the cluster for selected epochs for the sloshing simulations
including cooling and conduction. Anisotropic conduction, even in the case where it is most
efficient, is unable on its own to suppress the runaway cooling in the
very center of the cluster, with central temperatures falling to the
temperature floor of our simulation ($T \sim 0.01$~keV), far below the
temperatures observed in real clusters. However, the temperature is
significantly raised outside the very central region of the cool core, at radii $r \simgt 10$~kpc. 

\begin{figure*}
\begin{center}
\includegraphics[width=0.37\textwidth]{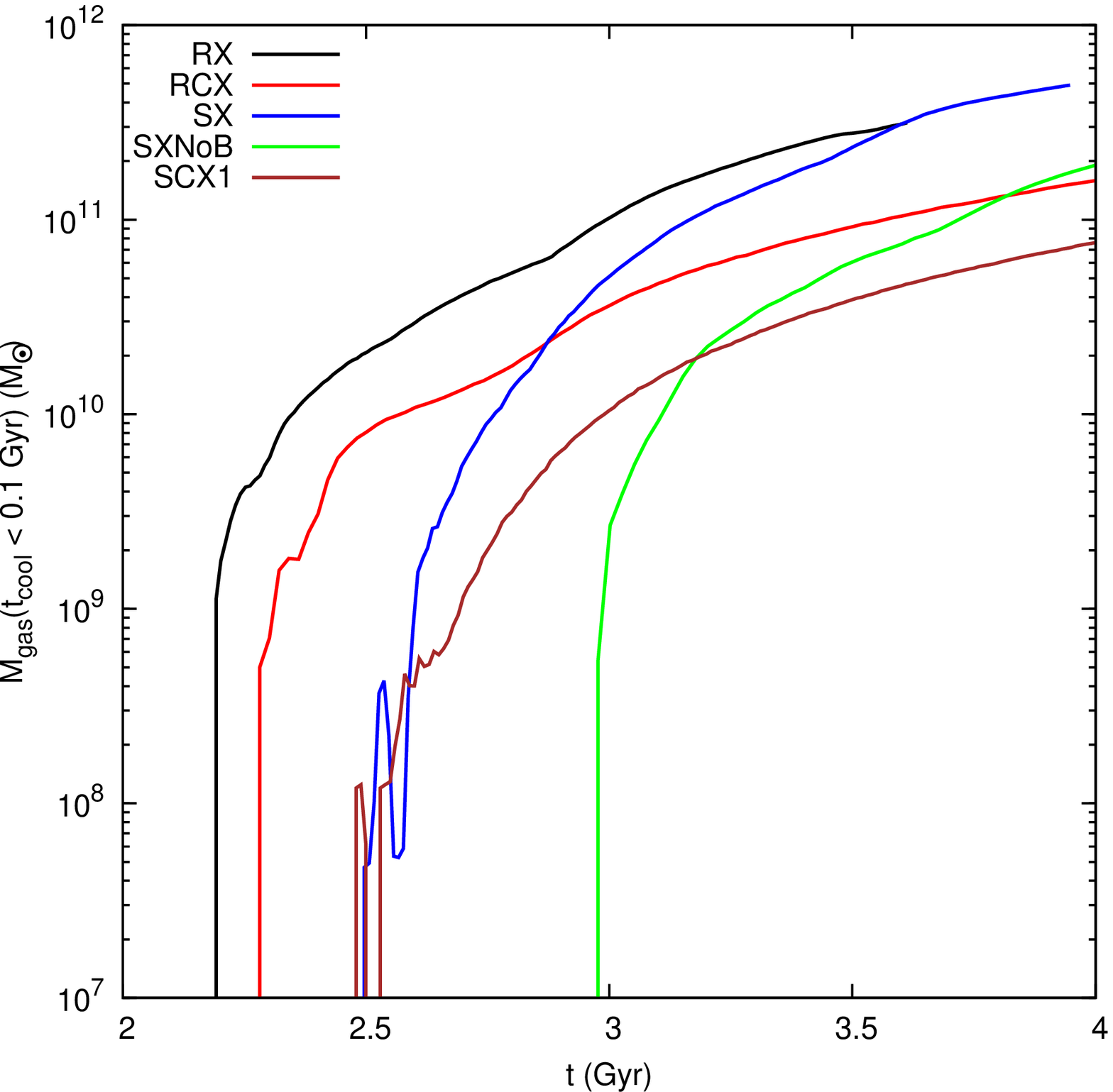}
\includegraphics[width=0.37\textwidth]{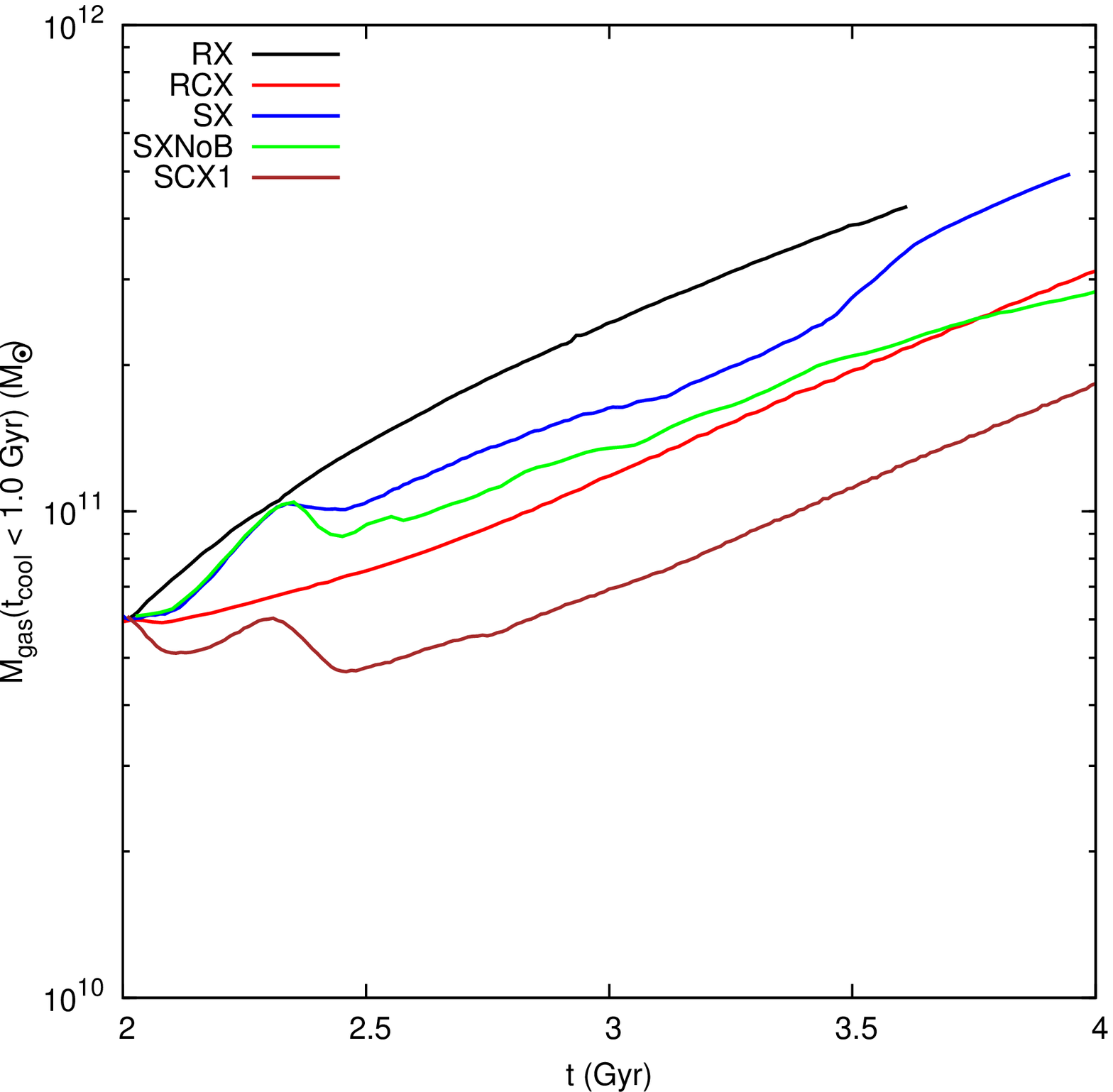}
\enspace
\includegraphics[width=0.8\textwidth]{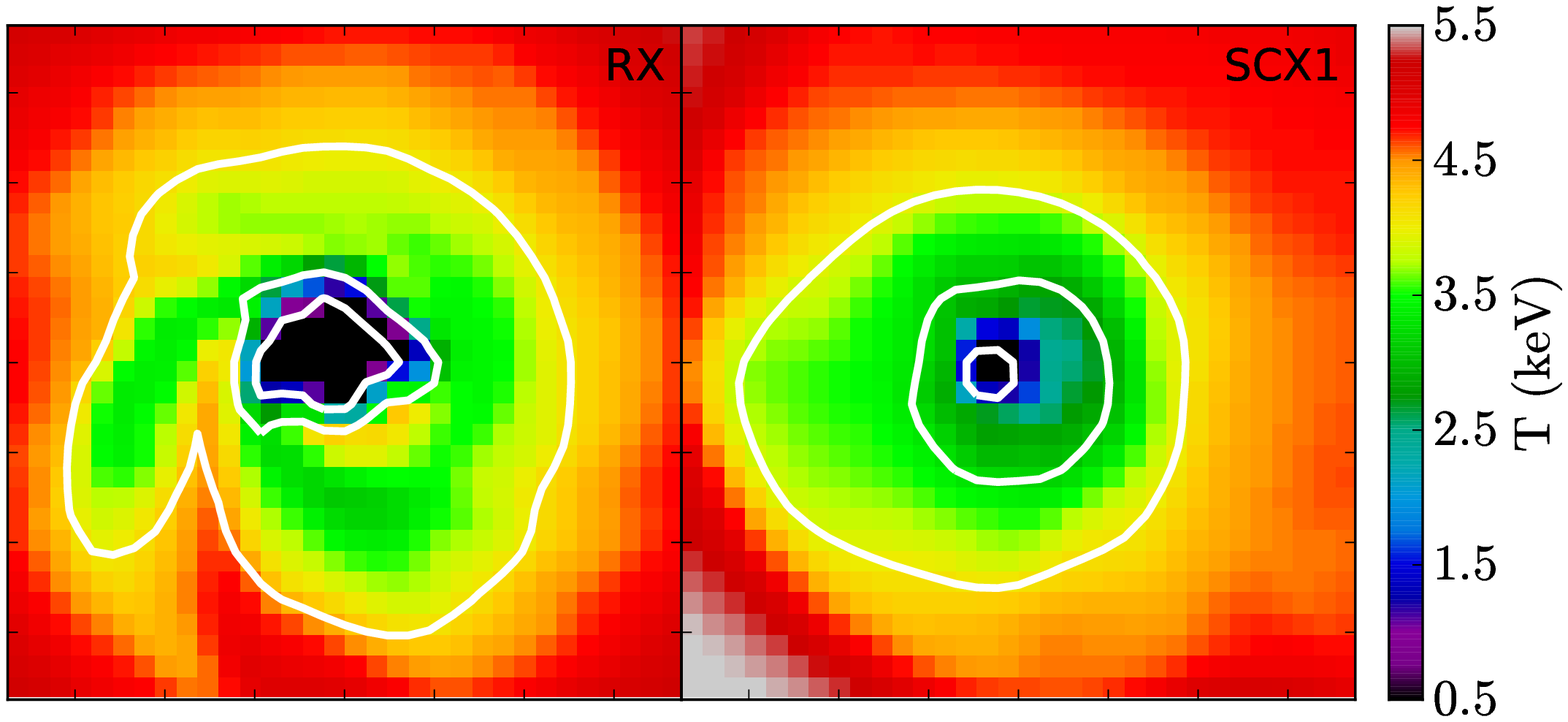}
\caption{Top panels: Mass of gas with a cooling time below a given
  value vs. simulation time for several of the cooling
  simulations. Left panel: Mass of gas with $t_{\rm cool} <
  0.1$~Gyr. Right panel: Mass of gas with $t_{\rm cool} <
  1.0$~Gyr. Bottom panels: Slices through the $z = 0$ plane
  of the temperature in keV for the {\it RX} and {\it SCX1} simulations at
  the epoch $t$ = 3.5~Gyr. White contours mark the surfaces with
  $t_{\rm cool}$ = 1.0, 0.5, and 0.1~Gyr. Each panel is 75~kpc on
  a side. Tick marks indicate 10~kpc distances.\label{fig:cool_mass}}
\end{center}
\end{figure*}

The two leftmost panels of Figure \ref{fig:cool_mass} show the evolution of the
gas mass with short cooling times within the inner
50~kpc of the cluster center over time. We were only able
to run the {\it RX} simulation out to an epoch of $t \sim 3.6$~Gyr, as
the extremely high densities, low temperatures, and magnetic pressures
in this simulation made the simulation too unstable to continue past
this point. Finally, we have also included a cooling run {\it SXNoB}, identical to
simulation {\it SX}, but without magnetic fields, to compare the
rate of increase of cool gas when the mixing of hot and cool gases is not
suppressed by such fields. This the same simulation as in ZMJ10, but
the setup (resolution and model for the gravitational potential) is
identical to our present work, to enable a direct comparison.

The top panels of Figure \ref{fig:cool_mass} show the evolution of the
gas mass with short cooling times within the inner
50~kpc of the cluster center over time. We were only able
to run the {\it RX} simulation out to an epoch of $t \sim 3.6$~Gyr, as
the extremely high densities, low temperatures, and magnetic pressures
in this simulation made the simulation too unstable to continue past
this point. Finally, we have also included a cooling run {\it SXNoB}, identical to
simulation {\it SX}, but without magnetic fields, to compare the
rate of increase of cool gas when the mixing of hot and cool gases is not
suppressed by such fields. This the same simulation as in ZMJ10, but
the setup (resolution and model for the gravitational potential) is
identical to our present work, to enable a direct comparison.

Over the course of $\sim$2~Gyr, all of the simulations have a
significant increase of the mass of cool gas in
their cores. However, both sloshing and conduction reduce the buildup
of cool gas. The top-left panel of Figure \ref{fig:cool_mass} shows
the buildup of gas mass with $t_{\rm cool} < 0.1$~Gyr. At the onset of
cooling, there is no gas in the core with a cooling time this short. Within $\sim$0.25~Gyr, in the relaxed
cluster simulations ({\it RX} and {\it RCX}), a gas mass of
$\sim~10^9-10^{10}~M_\odot$ with such cooling time has built
up in the core. This is somewhat slowed by conduction in the {\it RCX}
simulation, and by $t \sim$~4~Gyr the mass of cool gas is lower than
that of the {\it RX} simulation by a factor of a few. In the
simulations with sloshing ({\it SX} and {\it SCX1}), the initial
buildup of cool gas does not begin until $\sim$0.5~Gyr after the onset
of cooling. However, in the {\it SX} simulation, the mass of cool gas
eventually catches up with that of the {\it RX} simulation. Out of all
of these simulations, the run with both sloshing and conduction ({\it
  SCX1}) is most effective at suppressing the cooling of the gas for
the long term, with nearly an order of magnitude less mass of cool gas in the core than in
the {\it RX} simulation.

In the case of simulation {\it SXNoB}, no gas below
$t_{\rm cool} = 0.1$~Gyr appears until $\sim$1~Gyr after the onset of
cooling. Gas mixing, which is much more efficient in the absence of a
magnetic field (or perhaps in the presence of field variations
on microscopic scales that would strongly reduce field tension), provides a source of heat to the core gas and
prevents the catastrophic cooling of gas until this later
stage. However, the mass of cool gas increases quickly after this
point and is eventually comparable with that of simulation {\it
  RCX}. This indicates that in the absence of magnetic field tension, strong gas
mixing due to sloshing is a comparable source of heat to anisotropic
conduction. 

The top-right panel of Figure \ref{fig:cool_mass} shows
the buildup of gas mass with $t_{\rm cool} < 1.0$~Gyr. In this case,
all simulations already start with the same mass of gas with cooling
timescales this short. However, the general trend for each of these
simulations is the same as for the case for the gas mass with $t_{\rm
  cool} < 0.1$~Gyr. The runs with conduction (or gas mixing) and sloshing exhibit a
slower buildup of cool gas. 

The bottom panels of Figure \ref{fig:cool_mass} show slices through the
$z = 0$ plane of the temperature at the epoch $t$ = 3.5~Gyr, with
contours marking the surfaces with specific cooling times of 0.1, 0.5, and 1.0~Gyr, for the simulations {\it RX} and {\it SCX1}. It is
clear that the effect of sloshing and conduction together is to reduce
the volume of the region with very low cooling times and temperatures. Without
conduction and sloshing, a $r \sim 10$~kpc region with $t_{\rm cool} <
0.1$~Gyr develops, whereas with both of these effects included, this
region is restricted to $r \sim 3$~kpc. 

Overall, this result indicates that sloshing, even with anisotropic thermal
conduction, is not enough to completely stabilize a cool core against
a cooling catastrophe, and that for the central regions of clusters,
AGN feedback or other mechanisms are necessary to regulate the temperature of the
core. However, sloshing and conduction are able to suppress
or significantly delay catastrophic cooling in the cluster cooling
cores ($r \sim 50-100$~kpc) outside the very central regions of
radii $r \simlt 3$~kpc.

\section{Discussion and Conclusions\label{sec:conc}}

Gas sloshing, a prevalent phenomenon in relaxed galaxy clusters, is
evidenced by the nearly ubiquitous presence of spiral-shaped cold
fronts in cluster cool cores. The temperature jumps across these
fronts imply that either thermal conduction is
intrinsically weak in galaxy clusters or the magnetic field is
oriented parallel to the front surfaces, restricting conduction across
the fronts. Most previous works have assumed the
latter, and simulations have confirmed that such magnetic layers form
around cold fronts in clusters.  

Sloshing may also have an important role in determining the thermal evolution of the cluster
core. It brings the cold gas of the core into contact with hot
gas, possibly facilitating a transfer of heat
between these phases, either via mixing of the hot and cold gas or by
heat conduction. This raises the possibility that sloshing may be
partially responsible for preventing a ``cooling catastrophe'' in the
cores of galaxy clusters. However, the ICM is magnetized, which, as we
showed in ZML11, partially suppresses mixing. 

To determine whether anisotropic heat conduction in the
 presence of magnetic fields has any effect
on sloshing cold fronts and the thermal state of the cool cores, we
have performed MHD simulations with various configurations of the
magnetic field and the conduction coefficients. We determine
that, in spite of the fact that the cold fronts indeed are mostly draped by a
strong parallel magnetic field, they are not completely thermally
isolated from hotter gas. The shear
amplification and stretching of magnetic field lines occurs only in a
thin layer at some of the cold front surfaces, which only prevents heat
conduction to the cold gas below the front from the hot gas directly
above the front. Heat can still flow around the cold front, along field lines connected to
the cold gas from other regions, either from the hot gas at higher
radii perpendicular to the sloshing plane or from the hot flows that
are brought inside the cool core by sloshing. Thus, while the sharpness of the temperature
gradient at the cold front surfaces may be preserved at some locations
along the fronts, the cool gas below the fronts can still be heated
from other directions. Also, at some points along the cold front surfaces,
the sharpness of the temperature gradient is not preserved because a
stabilizing parallel magnetic field layer fails to form. Conduction also leads to
the reduction of the density jumps and the widening of the cold front interfaces, which can aid in
suppressing Kelvin-Helmholtz instabilities. Importantly, for the hot simulated cluster presented in this
work, we find that if conduction is unsuppressed along the field
lines, the density jumps can be smoothed to such a degree that the
cold fronts are no longer apparent in synthetic X-ray observations, in
profound disagreement with observed clusters. Due to the strong
temperature dependence of Spitzer conduction, we cannot use our
model cluster, which is hot ($T_X \sim 8$~keV), to draw conclusions
about the smearing out of cold fronts due to conduction in cooler
clusters. In a future paper, we will attempt more quantitative
comparisons with observed cold fronts in other clusters in order to
more generally constrain conduction. 

We find that a conduction coefficient parallel to the field lines an
order of magnitude smaller than the Spitzer value has a nearly insignificant effect on the temperature of the core and
the density and temperature jumps at the cold front surfaces,
producing cold fronts that are in agreement with observations. On the
other hand, adding even a small perpendicular conduction (0.01 of
Spitzer), that may, for example, mimic the effect of field
line reconnection, has an effect on the core specific entropy similar
to the case with unsuppressed Spitzer conduction along the field lines (because the
temperature gradients across the field lines are higher), and slightly
smoothes out the temperature gradients of the fronts over the course
of a few Gyr.

In the absence of radiative cooling, gas mixing and the conduction of heat between
the hot and cold phases in the cluster core raises the temperature and
the entropy of the core, and if left unchecked, would eliminate
the cool core entirely, leaving behind a high-entropy isothermal core.
When radiative cooling is taken into account, we find that, even for
full Spitzer conduction along the field lines, a cooling
catastrophe still occurs in the cluster center, though the effects of conduction
and sloshing together can significantly suppress the buildup of cool
gas in the core and restrict the cooling catastrophe to a small ($r
\simlt 5$~kpc) volume in the very center. The fact that conduction is unable to suppress cooling completely in the cluster center is somewhat inevitable, as for isobaric perturbations the cooling rate increases with lower temperature and the heating rate from conduction decreases \citep{kun11}. AGN feedback or other mechanisms are still necessary to prevent gas from cooling in these central regions. However, our results show that sloshing and conduction together can produce a significant amount of heat to offset some cooling in the cluster core.

We note that we have only considered one merger scenario and one initial $\beta$ for the
magnetic field. Though different choices in these parameter spaces are
not expected to change our main conclusions, from our previous works
we may make reasonable guesses about the effect of changing our
initial conditions. Certain merger setups result in more mixing of hot
and cold gases than others (ZMJ10), whereas a lower
initial $\beta$ results in stronger magnetic field layers at the cold
front surfaces and hence more suppression of K-H instabilities
(ZML11). Most importantly for our considerations, the strong
temperature dependence of Spitzer conduction makes it particularly
strong for our $T \sim 8$~keV simulated cluster. In order to use
observations of cold fronts in real clusters to put constraints on suppression of conduction, a wider parameter space of
initial cluster temperatures and plasma $\beta$ values would need to
be explored. Such a study, including a quantitative comparison with a
sample of observed cold fronts, will be presented in a future work. 

\acknowledgments
JAZ thanks Ian Parrish and Mikhail Medvedev for useful discussions and advice. Calculations were performed using the computational resources of the National
Institute for Computational Sciences at the University of
Tennessee and the Advanced Supercomputing Division at NASA/Ames
Research Center. Analysis of the simulation data was carried out using the
AMR analysis and visualization toolset yt \citep{tur11}, which is
available for download at \url{http://yt-project.org}. JAZ is
supported by the NASA Postdoctoral Program. MR acknowledges NSF grant
1008454. The software used in this work was in part developed by the DOE NNSA-ASC OASCR Flash Center at the University of Chicago.

\end{document}